\documentclass[aps,prd,preprint]{revtex4}%
\usepackage{amsfonts}
\usepackage{amsmath}
\usepackage{amssymb}
\usepackage{graphicx}%
\graphicspath{{figures_consistent_s2/}}

\begin{document}
\title{Response of the Moon to an anisotropic stochastic gravitational-wave background}
\author{Yanyue Gao$^{1}$}
\author{Hongbin Zhang$^{1}$}
\author{Xing Bian$^{2}$}
\author{Baocheng Zhang$^{1}$}
\email{zhangbaocheng@cug.edu.cn}
\affiliation{$^{1}$School of Mathematics and Physics, China University of Geosciences,
Wuhan 430074, China}
\affiliation{$^{2}$Institute of Mechanics, Chinese Academy of Sciences, Beijing 100190, China}

\begin{abstract}
Using a layered lunar model and elastodynamic theory, we investigate how the
angular structure of an anisotropic stochastic gravitational-wave background
appears in the lunar surface response. We compare a B20-inspired Galactic
foreground template, representing the large-scale angular structure of
unresolved Galactic sources, with a synthetic template containing three
localized hotspots. We project each sky distribution onto the Moon's five
quadrupolar response channels and combine their correlations with the lunar
normal-mode response to obtain frequency-resolved surface-response maps.
Although the Galactic template is visibly anisotropic, correlations between
distinct channels are weak when its reference axis is aligned with the lunar
frame; a representative reorientation increases their strength. The strongly
nonaxisymmetric hotspot template produces larger correlations and more
pronounced two-dimensional surface patterns. For the hotspot template at the
lower frequencies studied, changes in the balance between radial and
horizontal responses alter the surface morphology. These results connect
directional background models to predicted lunar displacement patterns and
provide a basis for future assessments of detector placement.

\end{abstract}
\maketitle

\section{Introduction}

In 1916, Einstein, using the weak-field approximation, predicted the existence
of gravitational waves (GWs)~\cite{AE1916}, which arise from the time
variation of the mass quadrupole moment of astrophysical bodies \cite{AE1918}.
In 2015, the Laser Interferometer Gravitational-Wave Observatory (LIGO) team
made the first direct detection of GWs with the GW150914 event~\cite{AAA2016},
thereby confirming their existence. Current and planned GW observations now
cover a broad frequency range. At high frequencies, the LIGO-Virgo-KAGRA
Collaboration has reported compact-binary observations in the $10$-$10^{3}%
\,$Hz band~\cite{AAA2023}. The proposed Einstein Telescope aims to extend the
accessible band down to a few hertz through underground siting and advanced
optical and interferometric technologies~\cite{PAA2010}. At lower frequencies,
space-based interferometers including LISA, TianQin, and Taiji will target the
mHz band with long interferometric baselines and drag-free control, enabling
observations of compact Galactic binaries, massive black-hole mergers, and
possible early-Universe signals such as cosmic
strings~\cite{AAB2017,LCD2016,HW2017}. Deep-space Doppler tracking provides a
complementary low-frequency probe by monitoring GW-induced fractional
frequency shifts in spacecraft radio links~\cite{JWA2006}. At still lower
frequencies, pulsar timing arrays search for stochastic signals in the
nanohertz band, motivated in particular by supermassive black-hole
binaries~\cite{CCG2021}, while primordial GWs may be constrained at ultralow
frequencies through the $B$-mode polarization of the cosmic microwave
background~\cite{AAA2020}. Together, these approaches span a broad frequency
range and provide complementary sensitivity to GW sources from galaxy
evolution to early-Universe physics.

A significant target across this multiband program is the stochastic
gravitational-wave background (SGWB), because it contains the cumulative
contribution of sources that are individually unresolved and may also carry
relic information from the early Universe. Measurements of, or upper limits
on, the SGWB therefore constrain both astrophysical source populations and
cosmological processes. Such a background can be generated by the incoherent
superposition of compact binaries and stellar remnants, as well as by
inflation, cosmic strings, or first-order phase transitions. Cosmological
backgrounds are often treated as isotropic to first approximation, whereas
realistic astrophysical foregrounds can be anisotropic, with a GW energy
density that depends on sky direction and can be described through a multipole
expansion~\cite{Cusin2017,Jenkins2018Formalism}. Anisotropies in backgrounds
formed by the superposition of compact binaries have been modeled to assess
their detectability with terrestrial detector networks
\cite{Jenkins2018Predictions,Capurri2021Intensity}. Related studies have also
considered space-based interferometer constellations~\cite{Capurri2023Space}.
In the mHz band relevant to LISA, the unresolved Galactic population of double
white dwarfs is expected to generate a bright and strongly anisotropic
Galactic foreground, with power concentrated toward the Galactic plane and bulge.

Template-based anisotropic analyses provide a practical way to study such
directional backgrounds. In this approach, the angular distribution of the
SGWB is represented by a specified sky pattern, or by a small family of
physically motivated patterns, rather than by a completely free sky map. This
reduces the dimensionality of the problem and allows particular source models
to be tested directly. For example, Criswell \textit{et al.} used an
adjustable Galactic disk-and-bulge template, denoted B20, to describe the
large-scale angular distribution of the unresolved double-white-dwarf
foreground~\cite{Criswell2025}. Searches for anisotropic backgrounds have also
been performed with LVK data, and recent reviews summarize the status of
isotropic and anisotropic SGWB searches across the GW
spectrum~\cite{Abbott2021Search,Renzini2022Stochastic}.

These developments motivate a complementary question: how does a directional
SGWB affect the response of an extended elastic body? A passing GW can excite
the Moon's normal modes and produce surface displacements whose spatial
pattern depends on the incident sky distribution. Surface instruments could
sample this pattern. The use of elastic bodies as GW antennas dates back to
Weber's early work~\cite{JW1960}. Dyson developed the seismic response of the
Earth to a GW~\cite{FJD1969}, and Ben-Menahem later studied GW excitation of
the Earth's normal modes~\cite{ABM1983}.

Apollo seismic measurements established an empirical record of lunar ground
motion~\cite{NLD1982}. The Apollo 17 Lunar Surface Gravimeter experiment, with
Weber among its investigators, was designed to use the Moon as an instrumented
antenna for GW detection~\cite{Giganti1973}. Apollo-era seismic records have
also been used to place limits on an isotropic SGWB in the decihertz
band~\cite{CH2014}. The Moon's low seismic-noise environment has motivated
newer detector concepts, including GLOC~\cite{JL2021} and the Lunar
Gravitational-Wave Antenna (LGWA)~\cite{HAA2021,ASSA2024}. Their proposed
frequency ranges include portions of the $0.1$-$1$\thinspace Hz interval; the
relevant band depends on the instrument design. Improved layered and
normal-mode models of the lunar response have also been
developed~\cite{YCZ2024,BH2024,MVL2025}. Yan \textit{et al.} derived
overlap-reduction functions for proposed networks of lunar seismometers,
quantifying the correlated response of detector pairs to an SGWB, and used
them to forecast network sensitivity~\cite{YCZZSW2024}.

In this paper, we use a radially stratified lunar model and elastodynamic
theory to study the surface response to anisotropic SGWBs. Building on the
statistical description of stochastic backgrounds by Allen and
Romano~\cite{AR1999}, we represent each directional background by a sky
template and project it onto the Moon's five quadrupolar response channels. We
compare a B20-inspired Galactic angular template with a synthetic
three-hotspot template to examine how broad and localized sky structures
appear in the lunar response. We present frequency-resolved maps of the lunar
surface response. These maps may inform future detector-placement studies that
include instrument performance and local noise.

The remainder of this paper is organized as follows. In Sec.~II, we combine
the GW description with the key equations governing the response of a radially
inhomogeneous elastic body; the standard Green-tensor and normal-mode
derivations are collected in Appendix~\ref{app:response-formalism}. In
Sect.~III, we compute the lunar response to an anisotropic SGWB and explain
how the sky template enters the response calculation. In Sec.~IV, we analyze
the coherent response to a monochromatic GW and its frequency evolution. In
Sec.~V, we calculate the lunar response to the Galactic foreground using the
B20-inspired template. In Sec.~VI, we construct a toy hotspot template to
study how different angular distributions affect the lunar response. Finally,
we summarize our conclusions in Sec.~VII.

\section{Gravitational-wave forcing and elastic response}

The plane-wave solution for GWs can be obtained by solving the Einstein field
equations in the linearized weak-field approximation. For a distant source,
the perturbation is sufficiently weak at the detector and the metric can be
written as
\begin{equation}
\boldsymbol{g}_{\mu\nu}=\boldsymbol{\eta}_{\mu\nu}+\boldsymbol{h}_{\mu\nu
},\quad\left\vert \boldsymbol{h}_{\mu\nu}\right\vert \ll1,
\end{equation}
where $\boldsymbol{g}_{\mu\nu}$ is the metric tensor of the (3+1)-dimensional
spacetime, $\boldsymbol{\eta}_{\mu\nu}=\mathrm{diag}(-1,1,1,1)$ is the
Minkowski metric of flat spacetime, and $\boldsymbol{h}_{\mu\nu}$ is the
weak-field perturbation produced by a distant source. Since GWs have only two
physical degrees of freedom, we use the transverse-traceless (TT) gauge, in
which the spatial metric perturbation can be written as
\begin{equation}
\boldsymbol{h}=\Re\left\{  h_{0}\boldsymbol{\epsilon}e^{i\left(  \omega
_{g}t-\boldsymbol{k}_{g}\cdot\boldsymbol{r}\right)  }\right\}  ,
\label{eq:monochromatic_wave}%
\end{equation}
where $\Re$ means taking the real part, $h_{0}$ is the amplitude of GWs,
$\boldsymbol{\epsilon}$ is the polarization tensor, $\boldsymbol{k}_{g}$ is
the GW wave vector, with $|\boldsymbol{k}_{g}|=\omega_{g}/c$ for GW angular
frequency $\omega_{g}$ and speed of light $c$. The selenocentric coordinate
system and the construction of $\boldsymbol{\epsilon}$ from the
incident-direction and polarization angles are specified in
Appendix~\ref{app:coordinates} and Fig.~\ref{Fig-a1}. We write the incident
direction as $\Omega=(e,\lambda)$ and denote the polarization angle by $\nu$;
$(\theta,\varphi)$ specify a lunar observation point. In the single-mode
expression below, the incident-wave parameters are held fixed and omitted from
the displacement arguments.

Under the weak-field approximation, the elastodynamic equation of an elastic
body perturbed by GWs can be written in the TT gauge as
\begin{equation}
\rho\frac{\partial^{2}\boldsymbol{u}}{\partial t^{2}}=\mu\nabla^{2}%
\boldsymbol{u}+(\lambda_{l}+\mu)\nabla(\nabla\cdot\boldsymbol{u}%
)-\frac{\partial\mu}{\partial x^{j}}h^{ij},\label{eegw}%
\end{equation}
which is the Navier-Lam\'{e} equation. Here, $\lambda_{l}$ denotes the first
Lam\'{e} parameter and $\mu$ is the second Lam\'{e} parameter. The last term,
$-\frac{\partial\mu}{\partial x^{j}}h^{ij}$, is the effective body-force
density induced by the coupling between the spatial gradient of the shear
modulus and the metric perturbation \cite{FJD1969,ABM1983}. For any radially
stratified elastic body with $\mu=\mu(r)$, this force density becomes
\begin{equation}
\boldsymbol{f}(\boldsymbol{r},t)=-\frac{d\mu}{dr}\hat{\boldsymbol{e}}_{r}%
\cdot\boldsymbol{h}=\Re\left\{  -h_{0}\frac{d\mu}{dr}\hat{\boldsymbol{e}}%
_{r}\cdot\boldsymbol{\epsilon}e^{i(\omega_{g}t-\boldsymbol{k}_{g}%
\cdot\boldsymbol{r})}\right\}  ,
\end{equation}
with the lunar case obtained by inserting the lunar radial profiles for
$\rho(r)$ and $\mu(r)$.

The solution of Eq.~(\ref{eegw}) using the Green tensor and normal modes
follows standard treatments in global seismology and GW-elastic-body coupling
theory \cite{JW1960,FJD1969,ABM1983,DT2020,MRR2019}; we summarize the
derivation in Appendix~\ref{app:response-formalism}. Here we retain only the
result needed to connect the elastic response to an anisotropic SGWB sky
template. We use a nonrotating, spherically symmetric lunar model. We assume
the long-wavelength condition $k_{g}R\ll1$, under which the GW forcing is
approximately coherent across the Moon. We do not consider toroidal modes. For
this model, the angular projection selects only $l=2$ spheroidal normal modes
\cite{JW1960,FJD1969,ABM1983,BCC1996}.

Let $k=(n,l,m)$ label a normal mode, with displacement eigenfunction
$\boldsymbol{s}_{k}$. For $l=2$, $F^{m}(\Omega,\nu)=\epsilon_{ij}(\Omega
,\nu)\Lambda_{ij}^{(m)}$ projects the incident direction and polarization onto
the five lunar quadrupolar channels. The free-surface traction is
$\boldsymbol{\tau}=\mu(R)\hat{\boldsymbol{e}}_{r}\cdot\boldsymbol{h}$ at
$r=R$. The volume force and surface traction together determine the radial
coupling coefficient $\alpha_{n}$, which includes the normal-mode mass
normalization. The source projection and radial overlap are derived in
Appendix~\ref{app:angular-reduction}; the complete product $\alpha
_{n}\boldsymbol{s}_{n2m}$ is independent of the arbitrary overall
eigenfunction normalization. The damped normal-mode source function
$D_{n}(\omega_{g})$ is defined in Appendix~\ref{app:transfer-functions}. The
physical displacement is obtained by taking the real part after the complete
complex response has been formed. For a single monochromatic $l=2$ normal
mode, the result used below is
\begin{equation}
\boldsymbol{u}_{n2m}(\boldsymbol{r},t) =h_{0}\Re\left\{  e^{i\omega_{g}
t}D_{n}(\omega_{g})\alpha_{n} F^{m}(\Omega,\nu)\boldsymbol{s}_{n2m}%
(\boldsymbol{r})\right\}  . \label{rf2}%
\end{equation}

\section{Response to the SGWB}

In general, a SGWB can be generated by the random superposition of waves
emitted from different unresolved astrophysical sources, and it can also
contain contributions from cosmological processes in the early Universe. Its
statistical properties may be isotropic or anisotropic depending on the source
population and its angular distribution. While an isotropic SGWB is a useful
first approximation for many cosmological backgrounds, an anisotropic SGWB
contains directional information and can therefore probe the spatial
distribution of astrophysical foregrounds and source populations. The latter
case is the focus of this paper. The real stochastic TT perturbation is a
superposition of the monochromatic plane waves used in Sec.~II,
\begin{equation}
\begin{aligned} h_{ij}(\boldsymbol{r},t)=\sqrt{2}\,\Re\left\{ \sum_{A=+,\times}\int _{0}^{\infty}df\int_{S^{2}}d\Omega\,\tilde{h}_{A}(f,\Omega)e_{ij}^{A}(\Omega)e^{i[2\pi ft-\boldsymbol{k}_{f}(\Omega)\cdot\boldsymbol{r}]}\right\} \end{aligned},\label{eq:sgwb_superposition}%
\end{equation}
where $\boldsymbol{k}_{f}(\Omega)$ is the wave vector with the value
$\frac{2\pi f}{c}$, and $e_{ij}^{A}(\Omega)$ are the fixed plus and cross
polarization tensors. Each term has the form of the monochromatic wave in
Sec.~II and enters the body-force term of Eq.~(\ref{eegw}) by linearity. The
surface traction is also linear in $h_{ij}$ and superposes in the same way;
hence the normal-mode response of Sec.~II applies to every term. The factor
$\sqrt{2}$ fixes the positive-frequency convention for the one-sided spectrum
below. Its role in the displacement variance is shown in
Appendix~\ref{app:stochastic-response}. We assume a stationary, Gaussian, and
unpolarized background with uncorrelated incident directions. We further take
the frequency spectrum and angular distribution to be separable, with a
frequency-independent sky template $P(\Omega)$. The strain-amplitude
covariance is then written
as~\cite{AR1999,Cusin2017,Jenkins2018Formalism,Abbott2021Search,Renzini2022Stochastic}
\begin{equation}
\left\langle \tilde{h}_{A}(f,\Omega)\,\tilde{h}_{A^{\prime}}^{\ast}(f^{\prime
},\Omega^{\prime})\right\rangle =\delta_{AA^{\prime}}\,\delta(f-f^{\prime
})\,\delta^{2}(\Omega,\Omega^{\prime})\,S_{h}(f)\,P(\Omega
),\label{anisotropic_psd}%
\end{equation}
with $A\in\{+,\times\}$ and $f=\omega_{g}/(2\pi)>0$. Here $S_{h}(f)$ is the
one-sided strain power spectral density for each equally weighted
polarization, with units $\mathrm{Hz}^{-1}$; $\tilde{h}_{A}(f,\Omega)$ is the
positive-frequency random strain amplitude, and $P(\Omega)$ describes the
normalized angular distribution of GW power over the sky,
\begin{equation}
\int_{S^{2}}d\Omega\,P(\Omega)=1,
\end{equation}
so that the isotropic limit corresponds to $P(\Omega)=1/(4\pi)$. We use a
single notation $P(\Omega)$, with $P(e,\lambda)$ used when the two
incident-direction angles are written explicitly. For the reference-normalized
maps below we set $S_{h}(f)=1\,\mathrm{Hz}^{-1}$; an astrophysical prediction
requires the actual strain spectrum.

Equation~(\ref{anisotropic_psd}) enters the lunar response through the same
direction- and polarization-dependent factors $F^{m}(\Omega,\nu)$ introduced
in Sec.~II and defined explicitly in Appendix~\ref{app:angular-reduction}. For
a deterministic plane wave, $F^{m}$ weights the five quadrupolar response
components directly. For a stochastic background, the products $F^{m}%
F^{m^{\prime}\ast}$ must instead be averaged over the sky and polarization
angle using the angular power distribution $P(\Omega)$. Thus $S_{h}(f)$ fixes
the overall spectral amplitude of the background, whereas $P(\Omega)$
determines how the anisotropic sky couples the different lunar quadrupolar components.

In this form, the anisotropy is entirely encoded in the sky map $P(\Omega)$.
For isotropic backgrounds, no direction on the sky is preferred and one often
reports rotation-invariant response measures obtained after averaging over sky
direction and detector orientation. By contrast, for anisotropic backgrounds
the detector response depends explicitly on the relative orientation between
the detector and the angular power distribution $P(\Omega)$, and this
directional dependence must be retained. Since the true sky distribution is
generally not known in advance, it is useful to adopt physically motivated or
diagnostic templates. In this paper, we use two templates: a B20-inspired
Galactic foreground angular template and a synthetic hotspot template. The
Galactic template is used below to illustrate the calculation procedure; the
same formalism is applied to the hotspot template in later sections.

The B20 model motivates our angular template and provides a simple description
of the large-scale morphology of the Galactic foreground, with most of the
signal power concentrated toward the Galactic plane and
bulge~\cite{Boileau2021Spectral,Bartolo2022Probing,Criswell2025}. Although the
B20 model was introduced in the context of the LISA Galactic foreground in the
mHz band, its disk and bulge morphology can guide an angular approximation for
our purpose. Using this Galactic template as the angular weight in the sky
integral, we compute the corresponding lunar response after averaging over the
GW polarizations. The resulting response map is therefore determined jointly
by the intrinsic anisotropic structure of the Galactic foreground and the
directional sensitivity of the lunar quadrupolar response.

To connect the SGWB statistics to the elastic response of Sec.~II, we rewrite
the lunar surface displacement in a form that separates the lunar response
kernel from the GW angular factors. For a single deterministic plane wave,
summing Eq.~(\ref{rf2}) over the normal modes gives, at fixed frequency
$f=\omega_{g}/(2\pi)$,
\begin{equation}
\boldsymbol{u}_{\mathrm{coh}}(\theta,\varphi,t;f,\Omega,\nu)=\Re\left\{
h_{0}e^{i\omega_{g}t}\sum_{m=-2}^{2}\boldsymbol{\mathcal{G}}_{2m}%
(\theta,\varphi;f)\,F^{m}(\Omega,\nu)\right\}  .
\label{eq:coh_real_displacement}%
\end{equation}
Here the purely lunar part is collected into the complex $m$-resolved kernel
\begin{equation}
\boldsymbol{\mathcal{G}}_{2m}(\theta,\varphi;f):=A(\omega_{g})\,Y_{2m}%
\,\hat{\boldsymbol{e}}_{r}+B(\omega_{g})\left(  \partial_{\theta}Y_{2m}%
\,\hat{\boldsymbol{e}}_{\theta}+\frac{1}{\sin\theta}\partial_{\varphi}%
Y_{2m}\,\hat{\boldsymbol{e}}_{\varphi}\right)  , \label{eq:G2m_def}%
\end{equation}
where $A(\omega_{g})$ and $B(\omega_{g})$ are the radial and tangential
normal-mode transfer functions defined in Eqs.~(\ref{A}) and (\ref{B}) of the
Appendix, with $\omega_{g}=2\pi f$. Here and below, angular arguments of the
lunar spherical harmonics are suppressed: $Y_{lm}=Y_{lm}(\theta,\varphi)$.
This is the same response derived from Eq.~(\ref{rf2}), but it is rewritten so
that the lunar structure and normal-mode physics are contained in
$\boldsymbol{\mathcal{G}}_{2m}$, while all dependence on GW direction and
polarization is contained in $F^{m}(\Omega,\nu)$.

We define the strain-normalized complex response amplitude, distinct from the
real displacement above, by the $m$-resolved expansion
\begin{equation}
\boldsymbol{\mathcal{U}}_{\mathrm{coh}}(\theta,\varphi;f,\Omega,\nu
)=\sum_{m=-2}^{2}\boldsymbol{\mathcal{G}}_{2m}(\theta,\varphi;f)\,F^{m}%
(\Omega,\nu). \label{eq:uk_expand_m}%
\end{equation}
Thus $\boldsymbol{u}_{\mathrm{coh}}=\Re\{h_{0}e^{i\omega_{g} t}
\boldsymbol{\mathcal{U}}_{\mathrm{coh}}\}$; the real part is taken after the
complete normal-mode and quadrupolar sums. The transfer functions in
$\boldsymbol{\mathcal{G}}_{2m}$ are generally complex because of damping.

At fixed frequency, the normal-mode transfer functions in
$\boldsymbol{\mathcal{G}}_{2m}(\theta,\varphi;f)$ are evaluated at that
frequency, and the remaining stochastic average acts only on the angular and
polarization factors. For any function $X(\Omega,\nu)$ of sky direction and
polarization angle, we define the $P$-weighted average by
\begin{equation}
\langle X\rangle_{P}= \int_{S^{2}}d\Omega\,P(\Omega)\,\frac{1}{\pi}\int
_{0}^{\pi}d\nu\, X(\Omega,\nu). \label{eq:P_weighted_average}%
\end{equation}
Here $\langle\cdot\rangle_{P}$ denotes an averaging operation; a particular
average is evaluated once its argument $X(\Omega,\nu)$ has been specified.
This average generalizes the usual full-sky average to a genuinely anisotropic
background. At fixed frequency and surface position, the lunar kernels
$\boldsymbol{\mathcal{G}}_{2m}$ are held fixed and are not included in $X$.

The polarization-angle average is equivalent to the equal-power, independent
plus/cross statistics in Eq.~(\ref{anisotropic_psd}). In the fixed basis of
Eq.~(\ref{eq:sgwb_superposition}), set $F_{A}^{m}(\Omega)=e^{A}_{ij}%
(\Omega)\Lambda_{ij}^{(m)}$. The convention $\boldsymbol{\epsilon
}=\boldsymbol{e}_{+}+\boldsymbol{e}_{\times}$ in
Appendix~\ref{app:coordinates} gives $F^{m}=\mathcal{A}_{\nu}F_{+}%
^{m}+\mathcal{B}_{\nu}F_{\times}^{m}$, where $\mathcal{A}_{\nu}=\cos2\nu
-\sin2\nu$ and $\mathcal{B}_{\nu}=\cos2\nu+\sin2\nu$. Direct integration over
$0\leq\nu<\pi$ yields
\begin{equation}
\langle\mathcal{A}_{\nu}^{2}\rangle_{\nu}=\langle\mathcal{B}_{\nu}^{2}%
\rangle_{\nu}=1, \qquad\langle\mathcal{A}_{\nu}\mathcal{B}_{\nu}\rangle_{\nu
}=0, \quad\langle\cdot\rangle_{\nu}:=\frac1\pi\int_{0}^{\pi}d\nu\,(\cdot).
\label{eq:polarization_average}%
\end{equation}
Consequently $\langle F^{m} F^{m^{\prime}\ast}\rangle_{\nu}=\sum_{A} F_{A}^{m}
F_{A}^{m^{\prime}\ast}$, as required by the $\delta_{AA^{\prime}}$ in
Eq.~(\ref{anisotropic_psd}).

Appendix~\ref{app:stochastic-response} derives the response directly from
Eq.~(\ref{eq:sgwb_superposition}), by projecting its volume force and surface
traction onto the lunar normal modes. Retaining the random strain amplitudes
and summing the modes gives the positive-frequency surface-displacement
amplitude
\begin{equation}
\widetilde{\boldsymbol{u}}(\theta,\varphi;f) =\sum_{A}\int_{S^{2}}%
d\Omega\,\tilde h_{A}(f,\Omega) \sum_{m=-2}^{2}\boldsymbol{\mathcal{G}}%
_{2m}(\theta,\varphi;f) F_{A}^{m}(\Omega).
\label{eq:stochastic_response_amplitude}%
\end{equation}
The real stochastic displacement is $\boldsymbol{u}(\theta,\varphi
,t)=\sqrt2\Re\{\int_{0}^{\infty}df\, e^{i2\pi ft}\widetilde{\boldsymbol{u}%
}(\theta,\varphi;f)\}$.

For the calculation, we adopt a B20-inspired angular template based on the
disk and bulge density profiles of Ref.~\cite{Criswell2025}. Thus, the sky
template $P(\Omega)$ entering the lunar response is obtained by projecting the
three-dimensional Galactic source distribution along each line of sight. The
prescription below defines the aligned Galactic template used in the numerical
results. Figure~\ref{Fig4} shows its Galactic-plane and bulge structure. The
unnormalized angular template is written as
\begin{equation}
\begin{aligned} \widetilde{P}(\Omega)={}&\int_{s_{\min}}^{s_{\max}}ds\,W(s)\, \bigl[A_{\mathrm{disk}}\rho_{\mathrm{disk}}(\varpi(s),z(s)) M_{\mathrm{disk}}(s,\Omega)\\ &\hspace{2em}+A_{\mathrm{bulge}}\rho_{\mathrm{bulge}}(\varpi(s),r^{\prime}(s))\bigr], \end{aligned} \label{eq:Ptilde}%
\end{equation}
where $s$ is the line-of-sight distance, $W(s)=s^{-2}$ is the distance
weighting adopted here, $\rho_{\mathrm{disk}}$ and $\rho_{\mathrm{bulge}}$ are
dimensionless disk and bulge density-profile shapes, and $A_{\mathrm{disk}}$
and $A_{\mathrm{bulge}}$ set their relative weights. The profiles used in the
numerical template are
\[
\begin{aligned}
\rho_{\mathrm{disk}}(\varpi,z)
&=\exp\!\left(-\frac{\varpi}{r_h}-\frac{|z|}{z_h}\right),\\
\rho_{\mathrm{bulge}}(\varpi,r^{\prime})
&=\frac{\exp[-(\varpi/r_{\mathrm{cut}})^2]}
{(1+r^{\prime}/r_0)^{p_{\mathrm b}}},\qquad
r^{\prime}=\sqrt{\varpi^2+(z/q_{\mathrm b})^2}.
\end{aligned}
\]
Here $r_{h}=3.31\,\mathrm{kpc}$ and $z_{h}=0.9\,\mathrm{kpc}$ are the disk
radial scale length and vertical scale height, respectively. For the bulge,
$q_{\mathrm{b}}=0.5$ is the vertical flattening parameter, $p_{\mathrm{b}%
}=1.8$ is the profile exponent, $r_{0}=0.075\,\mathrm{kpc}$ is the inner scale
radius, and $r_{\mathrm{cut}}=2.1\,\mathrm{kpc}$ sets the Gaussian cutoff in
cylindrical radius $\varpi$. The line-of-sight coordinates are
\[
\begin{aligned}
x(s)&=R_\odot-s\sin e\cos\lambda,&
y(s)&=s\sin e\sin\lambda,\\
z(s)&=z_\odot+s\cos e,&
\varpi(s)&=\sqrt{x(s)^2+y(s)^2},
\end{aligned}
\]
where $R_{\odot}=8.2\,\mathrm{kpc}$ and $z_{\odot}=0$ specify the observer's
Galactic position. The normalized sky template is then defined by
\begin{equation}
P(\Omega)=\frac{\widetilde{P}(\Omega)}{\int_{S^{2}}d\Omega\,\widetilde
{P}(\Omega)}. \label{eq:P_from_Ptilde}%
\end{equation}

For the aligned template we set $A_{\mathrm{disk}}=1$ and $A_{\mathrm{bulge}%
}=20$, and use 300 geometrically spaced points over $0.05\leq s\leq
30\,\mathrm{kpc}$. These relative weights are phenomenological; the overall
angular power is fixed by the normalization in Eq.~(\ref{eq:P_from_Ptilde}).
The finite spatial mask is applied only to the disk term,
\begin{equation}
M_{\mathrm{disk}}(s,\Omega)=\Theta\!\left(  4r_{h}-\varpi(s)\right)
\,\Theta\!\left(  4z_{h}-|z(s)|\right)  ,
\end{equation}
where $\Theta$ is the Heaviside step function. The bulge retains its intrinsic
radial taper and is not multiplied by $M_{\mathrm{disk}}$.

Once the anisotropic sky template $P(\Omega)$ is specified, we apply
Eq.~(\ref{eq:P_weighted_average}) to the product $X(\Omega,\nu)=F^{m}%
(\Omega,\nu)F^{m^{\prime}\ast}(\Omega,\nu)$ to obtain the quadrupolar
covariance matrix
\begin{align}
C_{mm^{\prime}}  &  =\left\langle F^{m}(\Omega,\nu)\,F^{m^{\prime}\ast}%
(\Omega,\nu)\right\rangle _{P}\nonumber\\
&  =\int_{S^{2}}d\Omega\,P(\Omega)\,\frac{1}{\pi}\int_{0}^{\pi}d\nu
\;F^{m}(\Omega,\nu)\,F^{m^{\prime}\ast}(\Omega,\nu)\nonumber\\
&  =\int_{S^{2}}d\Omega\,P(\Omega) \sum_{A=+,\times}F_{A}^{m}(\Omega
)F_{A}^{m^{\prime}\ast}(\Omega). \label{eq:quadrupolar_covariance}%
\end{align}
Here $m,m^{\prime}=-2,-1,0,1,2$. The second line writes out the average
defined in Eq.~(\ref{eq:P_weighted_average}). The third line uses
Eq.~(\ref{eq:polarization_average}) to evaluate the polarization integral,
leaving only the sky integral. These are equivalent expressions for the same
covariance; the polarization average is taken only once. For the
frequency-independent template adopted here, $C_{mm^{\prime}}$ is also
independent of frequency. The integrated variables $\Omega$ and $\nu$ are
therefore not arguments of $C_{mm^{\prime}}$. For an anisotropic background,
the matrix $C_{mm^{\prime}}$ is generally not diagonal, because the sky
template $P(\Omega)$ couples different $m$ components of the quadrupolar response.

Substituting Eq.~(\ref{eq:stochastic_response_amplitude}) twice into the
two-frequency displacement covariance and using Eq.~(\ref{anisotropic_psd}),
$\delta_{AA^{\prime}}$ removes the cross-polarization terms and $\delta
^{2}(\Omega,\Omega^{\prime})$ collapses one sky integral. The remaining
angular factors are $C_{mm^{\prime}}$, giving
\begin{equation}
\begin{aligned} \left\langle\widetilde{\boldsymbol u}(\theta,\varphi;f)\cdot \widetilde{\boldsymbol u}^{\ast}(\theta,\varphi;f')\right\rangle ={}&\delta(f-f')S_h(f)\sum_{m,m'=-2}^{2} \boldsymbol{\mathcal G}_{2m}(\theta,\varphi;f)\cdot \boldsymbol{\mathcal G}^{\ast}_{2m'}(\theta,\varphi;f) C_{mm'}. \end{aligned}\label{eq:displacement_covariance}%
\end{equation}
The frequency delta removes one frequency integral when the covariance is used
to form the total mean-square surface displacement,
\begin{equation}
\left\langle |\boldsymbol{u}(\theta,\varphi,t)|^{2}\right\rangle =\int
_{0}^{\infty}df\,S_{h}(f)\sum_{m,m^{\prime}=-2}^{2}\boldsymbol{\mathcal{G}%
}_{2m}(\theta,\varphi;f)\cdot\boldsymbol{\mathcal{G}}_{2m^{\prime}}^{\ast
}(\theta,\varphi;f)C_{mm^{\prime}}.\label{eq:total_mean_square}%
\end{equation}
Let $\langle|\boldsymbol{u}|^{2}\rangle_{\lbrack0,f]}$ denote the integral in
Eq.~(\ref{eq:total_mean_square}) with upper limit $f$. The fixed-frequency
quantity is the mean-square displacement spectral density $S_{u}%
=d\langle|\boldsymbol{u}|^{2}\rangle_{\lbrack0,f]}/df$:
\begin{equation}
S_{u}(\theta,\varphi;f)=S_{h}(f)\sum_{m,m^{\prime}=-2}^{2}%
\boldsymbol{\mathcal{G}}_{2m}(\theta,\varphi;f)\cdot\boldsymbol{\mathcal{G}%
}_{2m^{\prime}}^{\ast}(\theta,\varphi;f)C_{mm^{\prime}}%
.\label{eq:uk_intensity_general_aniso}%
\end{equation}
It has units $\mathrm{m}^{2}\,\mathrm{Hz}^{-1}$. The reference value
$S_{h}(f)=1\,\mathrm{Hz}^{-1}$ specifies a normalized strain spectrum;
physical total mean-square displacements require the frequency integral in
Eq.~(\ref{eq:total_mean_square}) with an astrophysical $S_{h}(f)$.

Finally, to obtain a lunar response closer to observational reality, we use
the MINEOS software~\cite{MWF2011} to compute the quadrupolar radial
eigenfunctions $U_{n}$ and $V_{n}$ entering Eq.~(\ref{rs}) after setting
$k=(n,2,m)$, and the transfer functions in Eqs.~(\ref{A}) and (\ref{B}). The
MINEOS eigenfunctions are converted to the vector-spherical-harmonic
convention used in this paper; in particular, the $l=2$ tangential component
carries the factor $1/\sqrt{l(l+1)}=1/\sqrt{6}$. The arbitrary overall
normalization of each eigenfunction cancels between the normal-mode coupling
numerator and the normalization integral in Eq.~(\ref{rs}), so the products
$U_{n}\alpha_{n}$ and $V_{n}\alpha_{n}$ used in the quadrupolar response are
normalization-consistent. The lunar structural model follows the realistic
lunar model in Ref.~\cite{YCZ2024}, which contains a homogeneous core and a
low-velocity zone (LVZ) outside the core. To improve accuracy in the
high-frequency band, the layered lunar model used here was generated by
interpolation to include 28,501 radial grid points. The model is synthesized
from several published works~\cite{WLG2011,GGC2011} in order to build a
complete input file usable by MINEOS. Unless otherwise stated, the normal-mode
sum is truncated at $n=400$.

\section{Instantaneous lunar surface response and its frequency evolution}

We first consider the coherent response to a deterministic monochromatic plane
GW, in order to illustrate the frequency dependence of the lunar surface
pattern. In the general normal-mode notation, $k=(n,l,m)$ denotes the radial
order, angular degree, and azimuthal order. Since the GW tidal field excites
only quadrupolar spheroidal modes, we set $l=2$ in the following and use $n$
to label the remaining radial overtones.

\begin{figure}[ptb]
\centering
\includegraphics[width=0.95\linewidth]{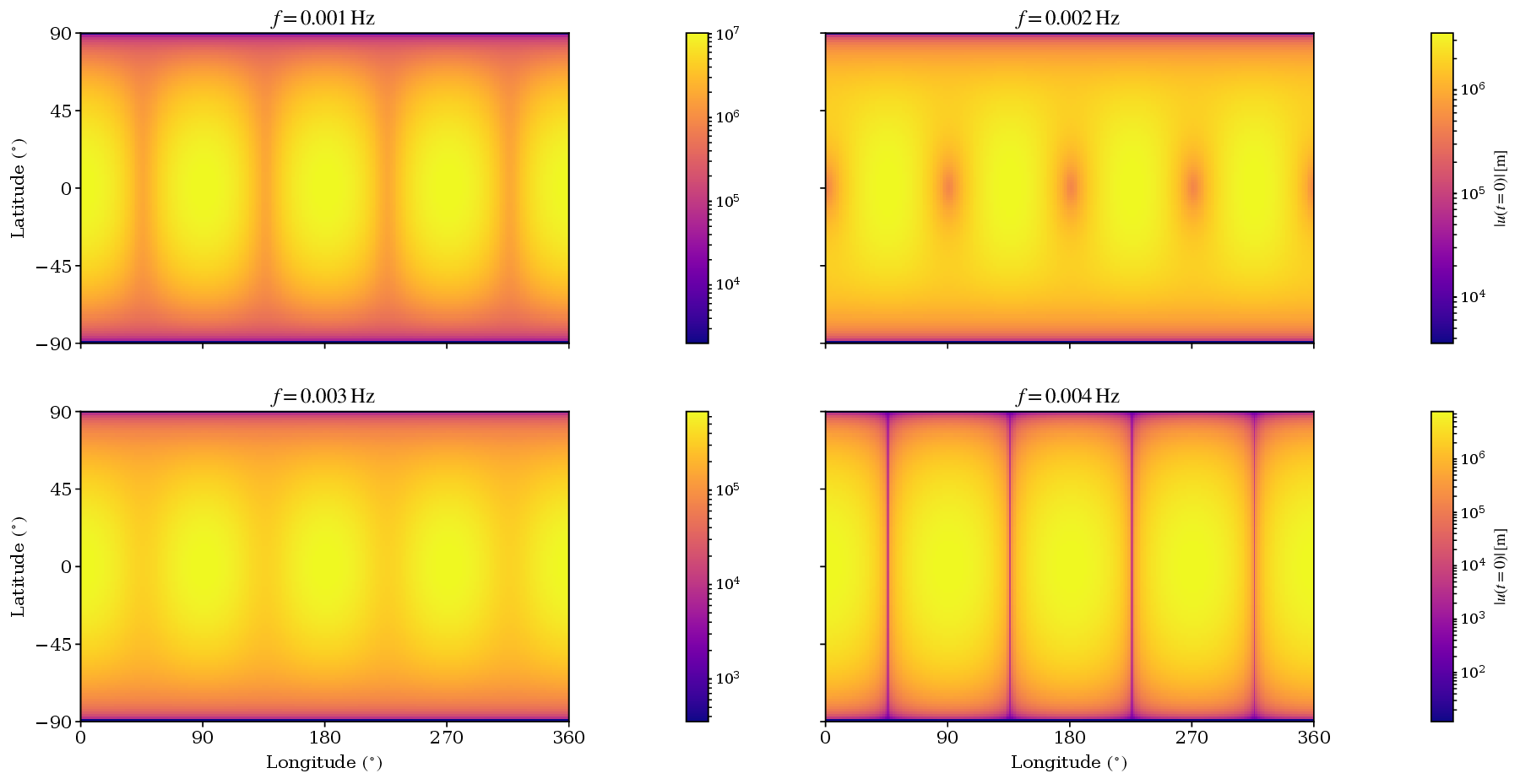} \caption{Magnitude of the
instantaneous real lunar surface displacement in
Eq.~(\ref{eq:coh_real_displacement}) for a monochromatic plane GW. The maps
use $h_{0}=1$, $t=0$, and $(e,\lambda,\nu)=(0,0,-\pi/8)$, giving
$F^{-2}=4\sqrt{2\pi/15}$ and all other $F^{m}=0$. They therefore show the
response per unit strain amplitude.}%
\label{Fig1}%
\end{figure}

For a GW incident along the $z$-axis, $e=0$. For a generic polarization, the
angular coefficients $F^{m}(\Omega,\nu)$ then satisfy
\begin{equation}
F^{0}=F^{\pm1}=0,\qquad F^{-2}\neq0,\qquad F^{2}\neq0 .
\end{equation}
Thus only the $m=\pm2$ quadrupolar channels can be driven. For the particular
polarization used in Fig.~\ref{Fig1}, $(e,\lambda,\nu)=(0,0,-\pi/8)$ and
$F^{-2}=4\sqrt{2\pi/15}$, while $F^{2}=0$. Here we use the real-harmonic
convention $Y_{2,-2}^{\mathrm{(r)}}=(Y_{2,2}^{\mathrm{(c)}}+Y_{2,-2}%
^{\mathrm{(c)}})/\sqrt{2}$, where the superscript $\mathrm{(c)}$ denotes
complex spherical harmonics; this gives the $\cos2\varphi$ angular combination
in the plotted maps. The maps show the magnitude of the real displacement in
Eq.~(\ref{eq:coh_real_displacement}). The quadratic form of the complex
coherent amplitude contains the angular products $F^{m}F^{m^{\prime}\ast}$.
Their sky and polarization average gives $C_{mm^{\prime}}$, which enters the
SGWB displacement spectral density in Eq.~(\ref{eq:uk_intensity_general_aniso}%
). The instantaneous coherent maps are obtained by taking the real part of the
full amplitude at the specified time before evaluating its magnitude.

\begin{figure}[ptb]
\centering
\includegraphics[width=0.55\linewidth]{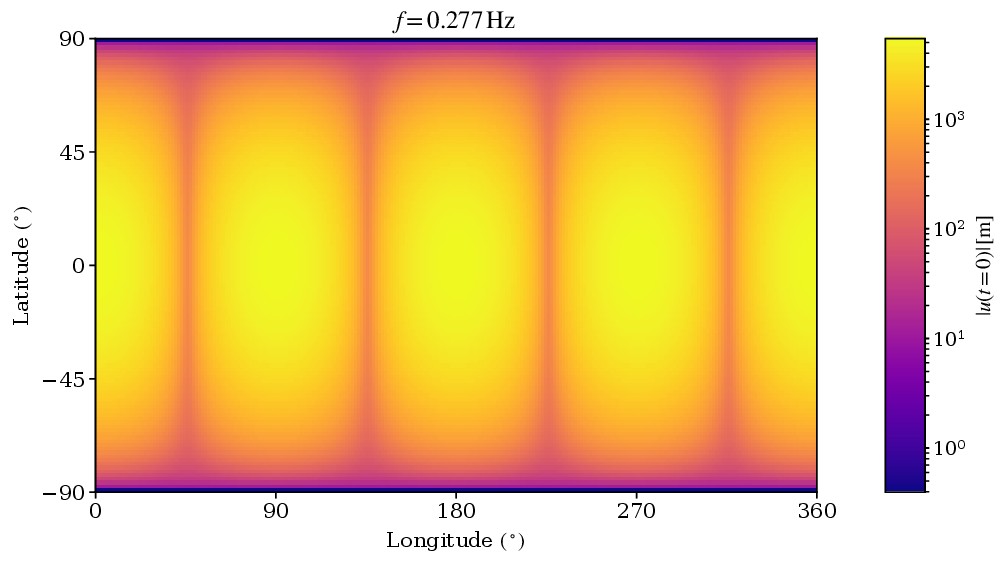} \caption{Instantaneous
coherent surface-response magnitude near $f\simeq0.277$~Hz, with
$(e,\lambda,\nu)=(0,0,-\pi/8)$, $h_{0}=1$, and $t=0$. This frequency lies
above the highest eigenfrequency included in the $n\le400$ normal-mode sum,
$f_{400}=0.273038$~Hz; the displayed pattern is therefore sensitive to
normal-mode truncation.}%
\label{Fig2}%
\end{figure}

For the deterministic maps in Figs.~\ref{Fig1} and \ref{Fig2}, no SGWB
spectral density is involved; we set $h_{0}=1$ to show the response per unit
monochromatic strain amplitude. The quantity $S_{h}(f)$ appears in the SGWB
statistical definition, Eq.~(\ref{anisotropic_psd}), and in the SGWB
displacement spectral density, Eq.~(\ref{eq:uk_intensity_general_aniso}). For
the SGWB maps, $S_{h}(f)=1\,\mathrm{Hz}^{-1}$ specifies the reference strain
spectrum, and the plotted quantity is a displacement spectral density in
$\mathrm{m}^{2}\,\mathrm{Hz}^{-1}$. We evaluate the instantaneous response at
$t=0$ to compare surface patterns across frequencies.

The qualitative change of the surface pattern near $f=0.002$~Hz in
Fig.~\ref{Fig1} follows from the frequency dependence of the radial and
horizontal transfer functions. We denote these complex functions by
$T_{r}(f)=A(2\pi f)$ and $T_{h}(f)=B(2\pi f)$, respectively, as defined in
Eqs.~(\ref{A}) and (\ref{B}). In the present spherically symmetric model, all
quadrupolar components share these functions. For the $t=0$ maps the relevant
coefficients are $\Re T_{r}$ and $\Re T_{h}$. At $f=0.002$~Hz, $|\Re
T_{r}|/|\Re T_{h}|=0.271$, so the horizontal contribution is stronger; at
$0.001$, $0.003$, and $0.004$~Hz the radial coefficient is stronger. This
change in the relative weight of $Y_{2m}$ and its angular derivatives shifts
the nodal and antinodal structure on the lunar surface. A mode-by-mode
decomposition of the horizontal transfer function shows that the horizontal
response near $0.002$~Hz is dominated by the $n=1$ mode, whose eigenfrequency
is $f_{n}=1.917866$~mHz.

\begin{figure}[ptb]
\centering
\includegraphics[width=1\linewidth]{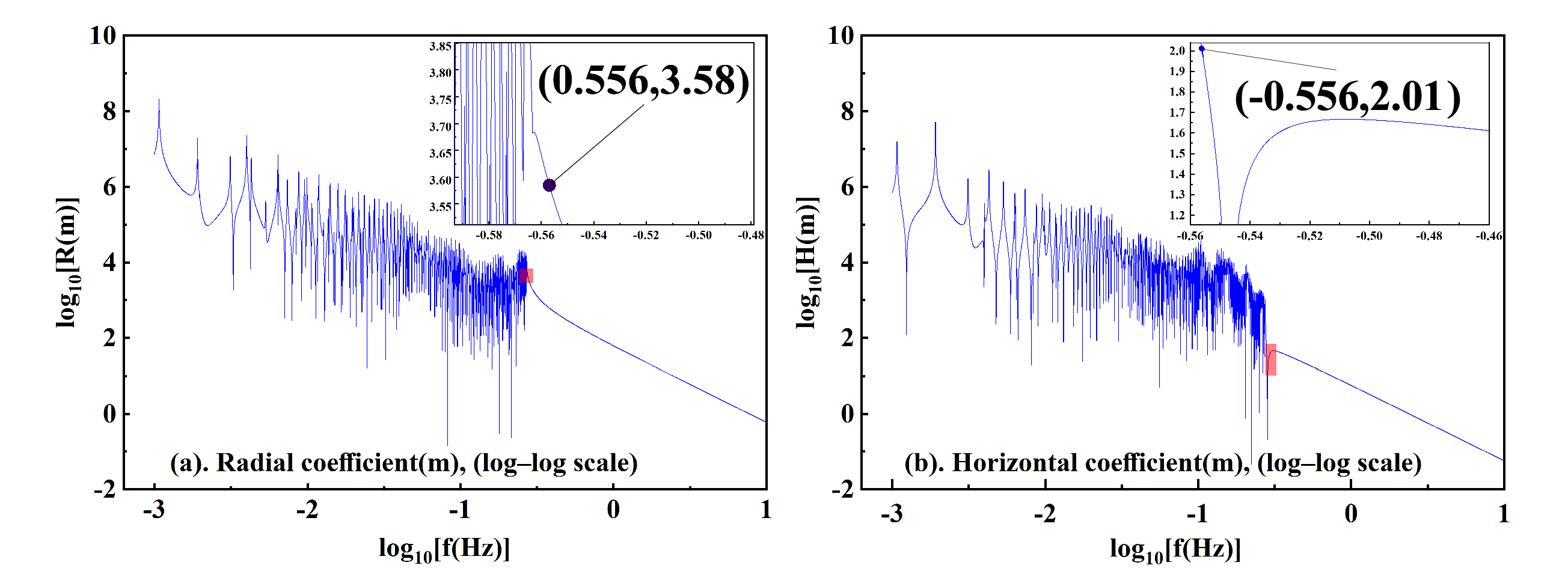} \caption{Fixed-phase radial and
horizontal transfer amplitudes as functions of frequency for the $n\le400$
normal-mode sum. The left and right ordinates show $\log_{10}[R(f)/\mathrm{m}%
]$ and $\log_{10}[H(f)/\mathrm{m}]$, respectively, versus $\log_{10}%
[f/\mathrm{Hz}]$. The reduced number of resolved peaks above $f_{400}%
=0.273038$~Hz reflects the finite normal-mode truncation.}%
\label{Fig3}%
\end{figure}

The normal-mode sum used for Figs.~\ref{Fig2} and \ref{Fig3} is truncated at
$n_{\max}=400$, with highest included eigenfrequency $f_{400}=0.273038$~Hz.
The smoother pattern shown near $0.277$~Hz lies just beyond this retained
spectrum and is therefore sensitive to the truncation. Including modes with
$n>400$ would extend the eigenfrequency coverage and could add resonant
features in this band. The displayed result therefore characterizes the
normal-mode sum truncated at $n=400$.

The fixed-phase transfer amplitudes shown in Fig.~\ref{Fig3} are
\begin{equation}
R(f)=|\Re T_{r}(f)|=|\Re A(\omega_{g})|,\qquad H(f)=|\Re T_{h}(f)|=|\Re
B(\omega_{g})|.
\end{equation}
Here $R(f)$ denotes the plotted radial transfer magnitude, distinct from the
lunar radius $R$ and the radial forcing overlap $\mathcal{R}_{n}$. At low
frequencies, nearby normal modes generate resonant peaks, so the coherent
surface pattern is sensitive to small changes in $f$. The detailed response
near the upper edge of the retained spectrum depends on the normal-mode
cutoff. Sufficiently far above all retained eigenfrequencies, each term of the
finite sum behaves approximately as
\begin{equation}
\frac{1}{\omega_{n}^{2}-\omega_{g}^{2}+i\omega_{n}\omega_{g}/Q_{n}}%
\simeq-\frac{1}{\omega_{g}^{2}},\qquad\omega_{g}\gg\omega_{n}.
\end{equation}
The truncated radial and horizontal transfer amplitudes then decrease with
frequency, as expected for the off-resonance limit of a finite normal-mode sum.

\section{Lunar response to the Galactic foreground}

\begin{figure}[ptb]
\centering
\includegraphics[width=0.8\linewidth]{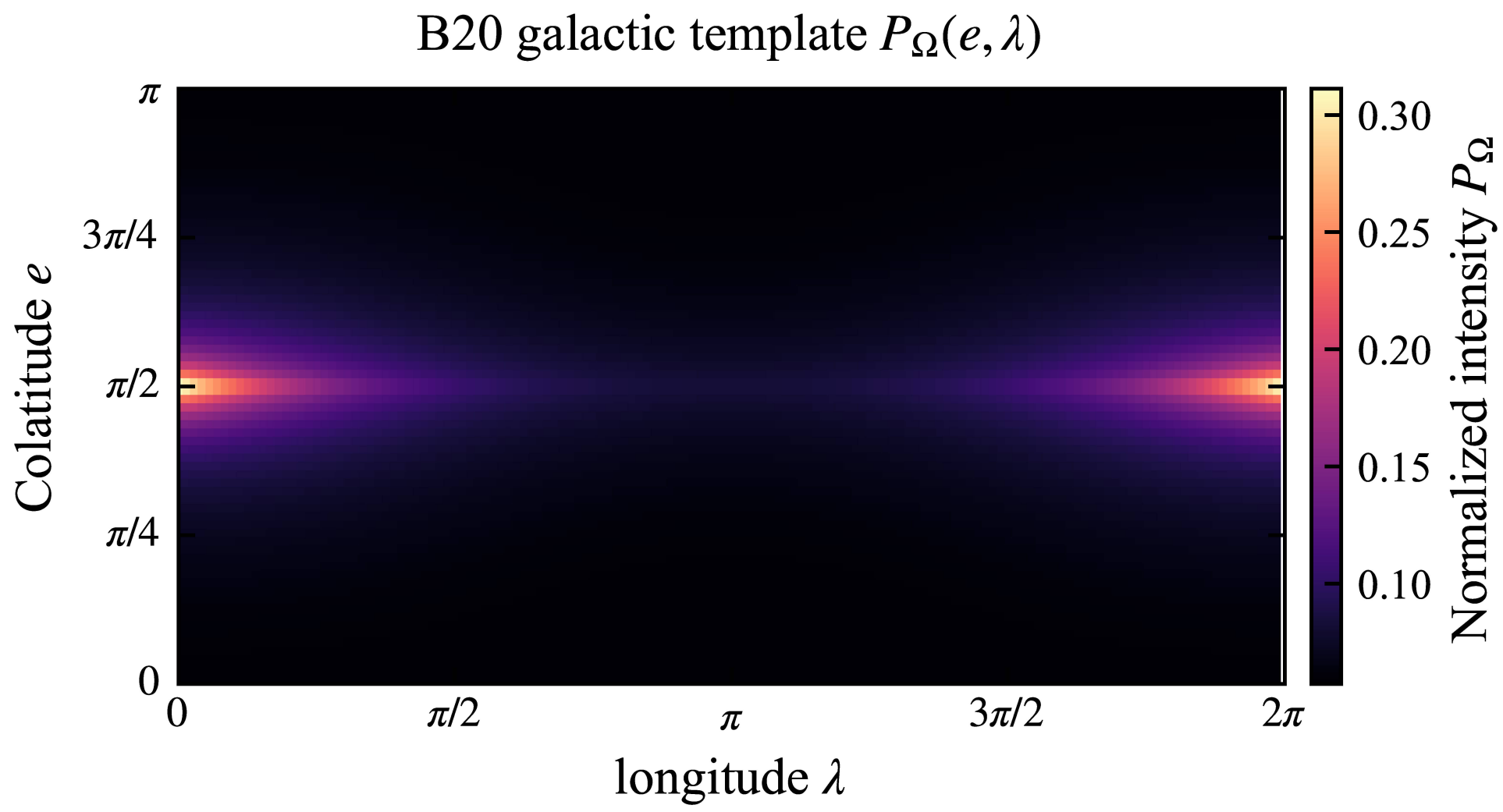}\caption{B20-inspired Galactic
angular template, normalized by $\int_{S^{2}}P(\Omega)\,d\Omega=1$. The
color-bar notation $P_{\Omega}$ denotes the normalized angular intensity
$P(\Omega)$. Power is enhanced near the Galactic plane ($e=\pi/2$) and toward
the bulge.}%
\label{Fig4}%
\end{figure}

Before presenting the lunar response, we first examine the angular structure
of the Galactic foreground adopted in this work. Figure~\ref{Fig4} shows the
B20-inspired Galactic template. The normalized $P(e,\lambda)$ used in the
response calculation is defined by Eqs.~(\ref{eq:Ptilde}) and
(\ref{eq:P_from_Ptilde}). Here $e$ and $\lambda$ denote the colatitude and
longitude on the sky, respectively. The template has enhanced power around
$e=\pi/2$, corresponding to the Galactic plane, and a weaker longitudinal
enhancement toward the central bulge near $\lambda\approx0$ and $2\pi$. Thus,
the sky distribution is nonuniform, although its departure from axisymmetry is
modest. In the following, we use the explicitly defined $P(\Omega)$ to compute
the lunar surface response for the cases in which the Galactic and lunar axes
are aligned or tilted relative to one another.

\subsection{The parallel case}

\begin{figure}[ptb]
\centering
\includegraphics[width=0.58\linewidth]{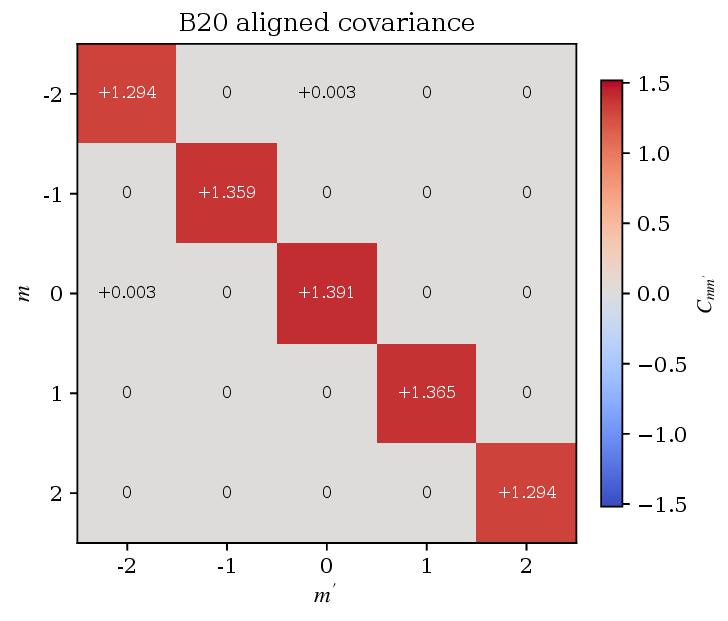}\caption{Projected covariance
matrix $C_{mm^{\prime}}$ for the aligned B20-inspired Galactic template,
calculated with tensor-normalized angular coefficients. Rows and columns are
ordered as $m,m^{\prime}=-2,-1,0,1,2$.}%
\label{Fig5}%
\end{figure}

Using the B20 template as the angular weighting function $P(\Omega)$, we
compute the quadrupolar covariance matrix,
\begin{equation}
C_{mm^{\prime}}=\int_{S^{2}}d\Omega\,P(\Omega)\,\frac{1}{\pi}\int_{0}^{\pi
}d\nu\,F^{m}(\Omega,\nu)\,F^{m^{\prime}\ast}(\Omega,\nu),\label{Cmm_B20}%
\end{equation}
where $m,m^{\prime}=-2,-1,0,1,2$. Numerically, $\int_{S^{2}}P(\Omega
)\,d\Omega=1$ to the accuracy of the angular quadrature. Because this angular
integral contains no normal-mode sum, $C_{mm^{\prime}}$ is independent of
$n_{\max}$; normal-mode convergence must instead be checked in the surface
response. Figure~\ref{Fig5} shows the aligned matrix. It is close to diagonal,
with its largest off-diagonal entry $C_{-2,0}=C_{0,-2}\simeq2.82\times10^{-3}%
$. The diagonal entries are nevertheless unequal, ranging from about $1.294$
to $1.391$. Thus, a nearly diagonal matrix does not imply an isotropic sky:
the aligned B20 template mainly changes the relative power of the five
quadrupolar channels. The isotropic control below separates that physical
effect from quadrature error.

\begin{figure}[ptb]
\centering
\includegraphics[width=0.95\linewidth]{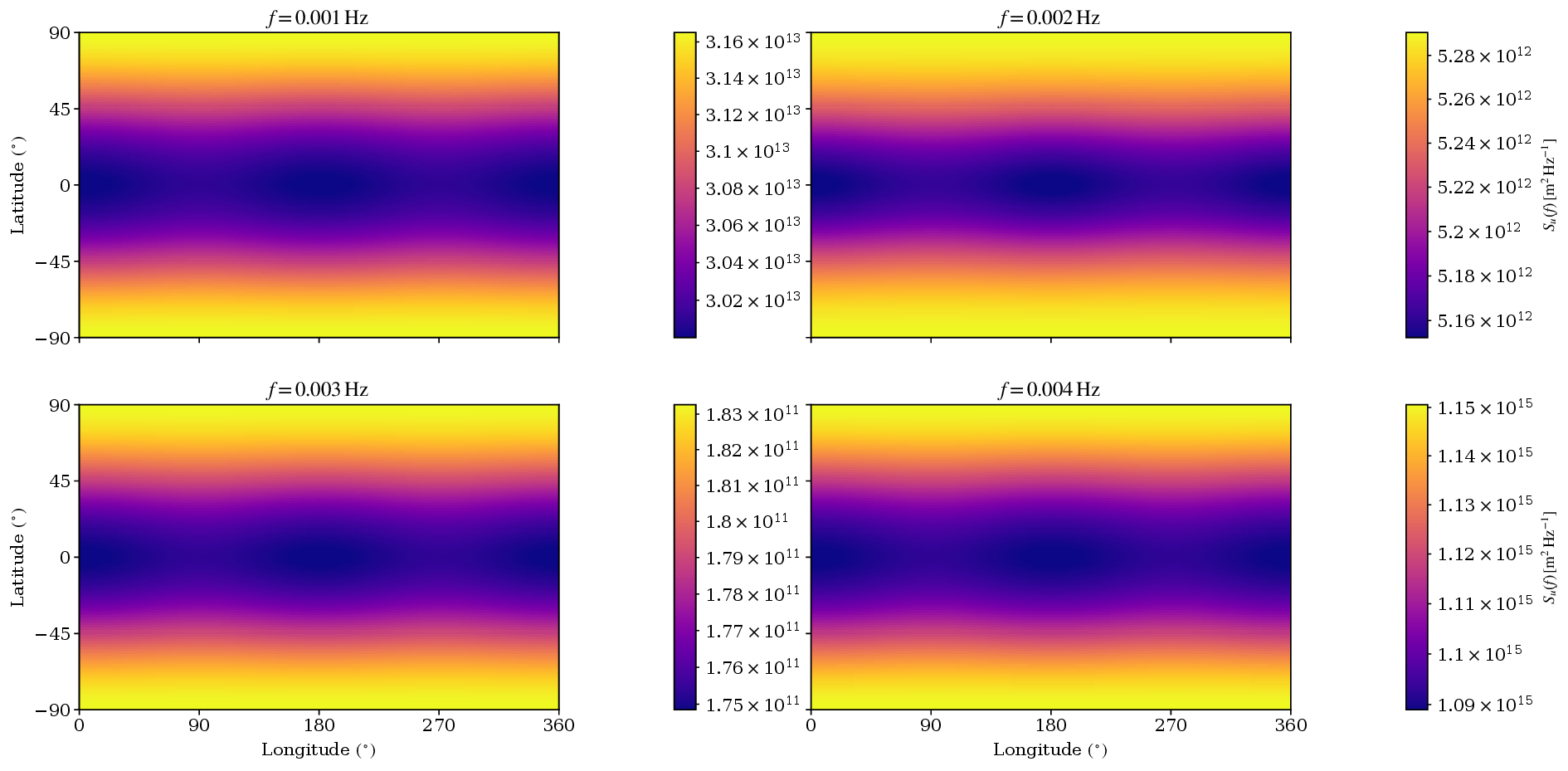}\caption{B20-weighted
mean-square displacement PSD $S_{u}(\theta,\varphi;f)$ at $f=0.001$, $0.002$,
$0.003$, and $0.004$~Hz, calculated with the complex transfer functions,
$n_{\max}=400$, and $S_{h}(f)=1~\mathrm{Hz}^{-1}$. Each panel gives a single
frequency and has units of $\mathrm{m}^{2}\mathrm{Hz}^{-1}$.}%
\label{Fig6}%
\end{figure}

\begin{figure}[ptb]
\centering
\includegraphics[width=0.55\linewidth]{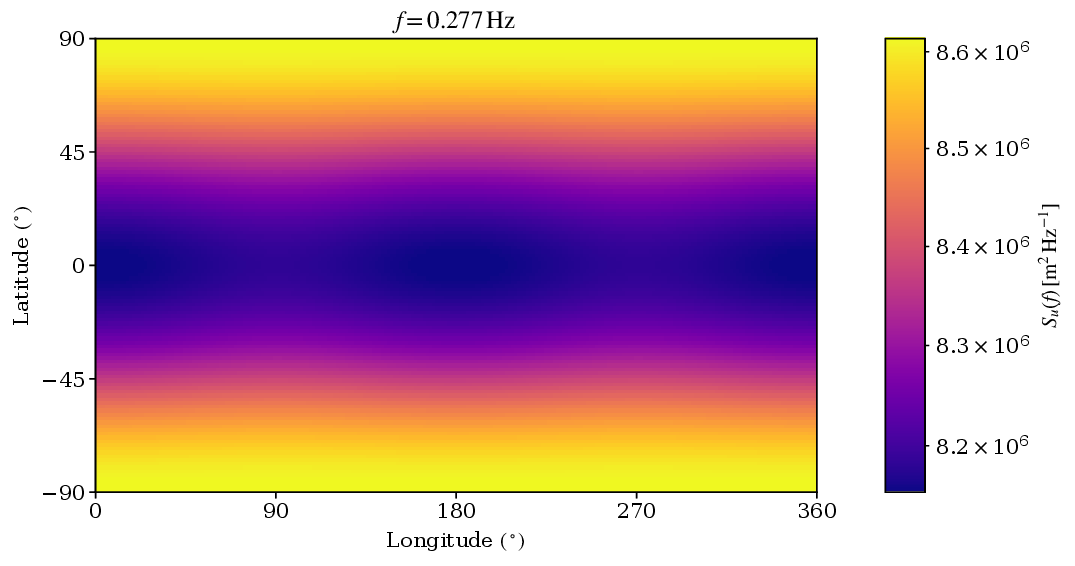}\caption{B20-weighted
mean-square displacement PSD at $f=0.277$~Hz, computed as in Fig.~\ref{Fig6}.
This frequency lies above the highest eigenfrequency in the $n\leq400$
normal-mode spectrum, and the absolute PSD is sensitive to the finite
normal-mode truncation.}%
\label{Fig7}%
\end{figure}

Substituting the covariance matrix given by Eq.~(\ref{Cmm_B20}) and shown in
Fig.~\ref{Fig5} into Eq.~(\ref{eq:uk_intensity_general_aniso}), we obtain the
mean-square SGWB spectral response. Figures~\ref{Fig6} and \ref{Fig7} plot the
displacement power spectral density $S_{u}(\theta,\varphi;f)$ for
$S_{h}(f)=1~\mathrm{Hz}^{-1}$, with units of $\mathrm{m}^{2}\mathrm{Hz}^{-1}$.
The calculation uses the long-wavelength approximation, retains only $l=2$
spheroidal modes, assumes a nonrotating spherically symmetric lunar model, and
truncates the radial-mode sum at $n_{\max}=400$. Each panel is evaluated at
one fixed GW frequency; no average over frequency is performed.

The B20 covariance matrix is independent of frequency. The frequency
dependence of the maps therefore enters through the radial and horizontal
transfer functions contained in $\boldsymbol{\mathcal{G}}_{2m}(\theta
,\varphi;f)$. In Figs.~\ref{Fig6} and \ref{Fig7}, the aligned-B20 response
retains essentially the same latitude-band pattern at all five displayed
frequencies. The color scales span only about $2$--$6\%$ of the mean value in
each panel, so the frequency dependence is expressed mainly through the
overall PSD amplitude. At $f=0.277$~Hz, the retained normal-mode spectrum ends
at $f_{400}=0.273038$~Hz. The $n\leq200$ and $n\leq400$ comparison below shows
that the absolute high-frequency PSD is not converged. Interpreting the
response physically above this range requires an extended mode spectrum.

Mathematically, the B20 template determines the angular covariance
$C_{mm^{\prime}}$, while the lunar kernels $\boldsymbol{\mathcal{G}}%
_{2m}(\theta,\varphi;f)$ determine the frequency and spatial dependence. The
full SGWB mean-square response follows from the double sum over $m,m^{\prime}$
in Eq.~(\ref{eq:uk_intensity_general_aniso}). The lunar response therefore
does not reproduce the Galactic sky map point-by-point.

\subsection{The tilted case}

To investigate how orientation affects the projected response, we tilt the
natural axis of the B20 template relative to the lunar reference axis. The
quadrupolar response basis is defined in the lunar frame. In the aligned case,
the template is approximately axisymmetric in that basis and the covariance is
nearly diagonal. Tilting the same sky distribution redistributes power among
the lunar-frame $m$ components and produces off-diagonal couplings.

We implement the tilt by an Euler rotation. Let
\begin{equation}
\mathbf{R}_{\mathrm{B20}\rightarrow\mathrm{moon}} =\mathbf{R}_{z}%
(\alpha)\mathbf{R}_{y}(\beta)\mathbf{R}_{z}(\gamma),
\end{equation}
where $\mathbf{R}_{z}$ and $\mathbf{R}_{y}$ are the usual right-handed
rotation matrices about the $z$ and $y$ axes:
\begin{equation}
\mathbf{R}_{z}(\psi)=
\begin{pmatrix}
\cos\psi & -\sin\psi & 0\\
\sin\psi & \cos\psi & 0\\
0 & 0 & 1
\end{pmatrix}
,\qquad\mathbf{R}_{y}(\vartheta)=
\begin{pmatrix}
\cos\vartheta & 0 & \sin\vartheta\\
0 & 1 & 0\\
-\sin\vartheta & 0 & \cos\vartheta
\end{pmatrix}
.
\end{equation}
A sky direction in the lunar frame, represented by $\hat{\boldsymbol{n}%
}_{\mathrm{moon}}(\Omega)$, is mapped into the intrinsic B20 frame by
\begin{equation}
\hat{\boldsymbol{n}}_{\mathrm{B20}} =\mathbf{R}_{\mathrm{moon}\rightarrow
\mathrm{B20}} \hat{\boldsymbol{n}}_{\mathrm{moon}},\qquad\mathbf{R}%
_{\mathrm{moon}\rightarrow\mathrm{B20}} =\mathbf{R}_{\mathrm{B20}%
\rightarrow\mathrm{moon}}^{T}.
\end{equation}
The rotated template used in the lunar-frame sky integral is therefore
\begin{equation}
P_{\mathrm{moon}}(\Omega)=P_{\mathrm{B20}}\!\left(  \mathbf{R}_{\mathrm{moon}%
\rightarrow\mathrm{B20}} \hat{\boldsymbol{n}}_{\mathrm{moon}}(\Omega)\right)
.
\end{equation}
In the numerical example we use $(\alpha,\beta,\gamma)=(50^{\circ},50^{\circ
},50^{\circ})$. Rotation preserves the normalization, and the numerical
evaluation gives $\int_{S^{2}}P_{\mathrm{moon}}(\Omega)d\Omega=1$ to numerical precision.

\begin{figure}[ptb]
\centering
\includegraphics[width=0.58\linewidth]{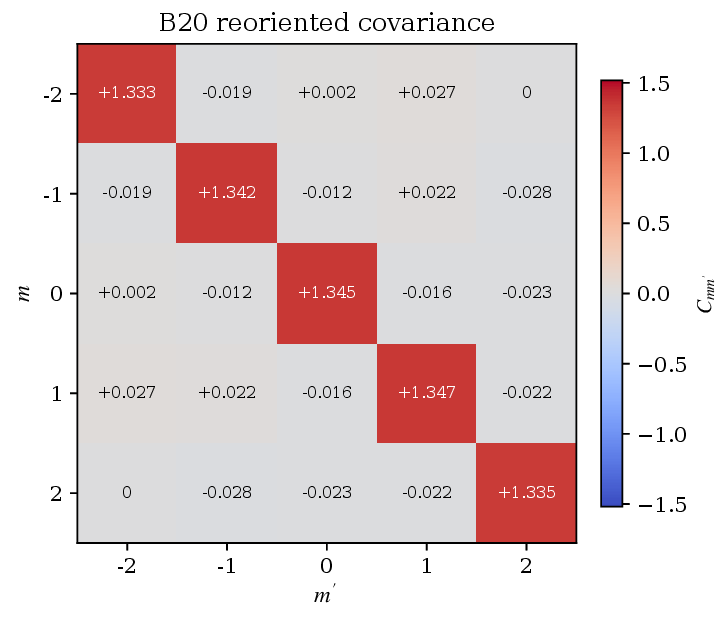}\caption{Projected covariance
matrix $C_{mm^{\prime}}$ for the tilted B20-inspired Galactic template,
calculated with tensor-normalized angular coefficients. Positive and negative
off-diagonal entries reveal stronger interchannel mixing than in the aligned
case.}%
\label{Fig8}%
\end{figure}

Using the tilted template in Eq.~(\ref{Cmm_B20}) gives the covariance matrix
in Fig.~\ref{Fig8}. Its off-diagonal entries increase, while the diagonal
terms remain dominant: $C_{-2,1}\simeq0.02689$, $C_{-1,1}\simeq0.02233$, and
$C_{-1,2}\simeq-0.02835$, for example. These off-diagonal entries are larger
than those in the aligned matrix, reflecting the stronger channel mixing in
the chosen lunar basis after reorientation. Rotation changes the channel
representation but preserves the covariance eigenvalues and its
rotation-invariant departure from a scalar identity.

To quantify this mixing in the chosen lunar frame, we define the off-diagonal
Frobenius-norm fraction
\begin{equation}
\eta_{\mathrm{aniso}}\equiv\frac{\Vert C-C^{(\mathrm{diag})}\Vert_{F}}{\Vert
C\Vert_{F}}=\left(  \frac{\sum_{m\neq m^{\prime}}|C_{mm^{\prime}}|^{2}}%
{\sum_{m,m^{\prime}}|C_{mm^{\prime}}|^{2}}\right)  ^{1/2},\label{eq:eta_aniso}%
\end{equation}
where $C_{mm^{\prime}}^{(\mathrm{diag})}=\delta_{mm^{\prime}}C_{mm}$. This
ratio measures interchannel coupling in the chosen frame, not the total
departure from isotropy; an anisotropic axisymmetric sky can have a diagonal
covariance matrix. The maximum possible value for a $5\times5$
positive-semidefinite covariance matrix is $\eta_{\max}=\sqrt{4/5}\simeq
0.894$. We obtain $\eta_{\mathrm{aniso}}\simeq1.33\times10^{-3}$ for the
aligned B20 template and $2.91\times10^{-2}$ for the tilted template. As a
complementary measure of the aligned matrix's total departure from an
isotropic covariance, define $c=\operatorname{tr}C/5$; then $\Vert
C-cI_{5}\Vert_{F}/\Vert C\Vert_{F}\simeq2.95\times10^{-2}$. The latter is
nonzero chiefly because the aligned diagonal entries are unequal. The tilted
case has stronger interchannel mixing; both matrices were computed with the
same angular routine that passes the isotropic control described below.

Here we compare the aligned and tilted cases at the covariance level in
Figs.~\ref{Fig5} and \ref{Fig8}. A tilted response map would follow by
substituting the matrix in Fig.~\ref{Fig8} into the same mean-square response
formula, Eq.~(\ref{eq:uk_intensity_general_aniso}); no additional formalism is required.

\subsection{Isotropic null test and normal-mode-cutoff sensitivity}

\label{sec:cutoff_sensitivity}

\begin{figure}[ptb]
\centering
\includegraphics[width=0.58\linewidth]{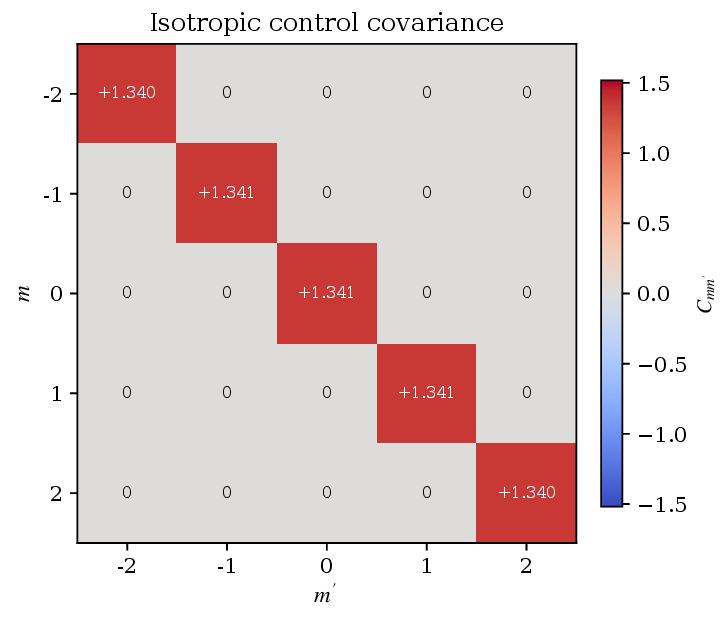}\caption{
Isotropic null-test covariance obtained with the same angular routine used for
Figs.~\ref{Fig5} and \ref{Fig8}. All five diagonal entries agree with
$32\pi/75$ to the accuracy of the angular quadrature; the off-diagonal entries
vanish numerically.}%
\label{FigIso}%
\end{figure}

To distinguish sky anisotropy from an angular implementation artifact, we
replace the B20 sky by $P_{\mathrm{iso}}(\Omega)=1/(4\pi)$ while retaining the
same tensor-normalized coefficient routine and sky quadrature. Rotational
invariance requires five equal diagonal entries and zero off-diagonal entries.
Analytically,
\begin{equation}
C_{\mathrm{iso}}=\frac{32\pi}{75}I_{5}. \label{eq:isotropic_corrected_check}%
\end{equation}
On the numerical grid the diagonal entries, in $m=-2,-1,0,1,2$ order, are
$1.340129$, $1.340602$, $1.340602$, $1.340602$, and $1.340129$, compared with
$32\pi/75\simeq1.340413$. Their small inequality is a quadrature effect: the
largest absolute difference from the analytic value is $2.84\times10^{-4}$,
while all off-diagonal entries are below $10^{-16}$. Using this finite-grid
covariance gives an isotropic surface PSD with relative spatial spread
$1.78\times10^{-4}$; substituting the analytic $C_{\mathrm{iso}}$ into the
same response code reduces the spread to $1.1\times10^{-15}$.
Figure~\ref{FigIso} shows the covariance control. Thus, the aligned B20
diagonal differences in Fig.~\ref{Fig5} exceed numerical quadrature error and
reflect the anisotropic sky template.

The angular covariance in Eq.~(\ref{Cmm_B20}) is independent of the
normal-mode cutoff $n_{\max}$. We therefore compare the mean-square PSD maps
$S_{u,200}$ and $S_{u,400}$ for the same aligned B20 covariance, using the
area-weighted relative difference $\delta=\Vert S_{u,200}-S_{u,400}%
\Vert_{L^{2}(S^{2})}/ \Vert S_{u,400}\Vert_{L^{2}(S^{2})}$ without separately
normalizing the two maps. We find $\delta=2.98\times10^{-4}$ at $0.002$~Hz and
$\delta=0.984$ at $0.277$~Hz. The low-frequency PSD is stable under this
change of cutoff, whereas the absolute high-frequency PSD is strongly
cutoff-dependent. Consequently, the $0.277$~Hz panel should be treated as a
truncated-model illustration rather than a converged prediction.

\section{The hotspot model}

In order to explore how different angular distributions of a SGWB affect the
lunar response, we now keep the general response formalism established above
and change only the sky template $P(\Omega)$. As a controlled example of a
strongly nonaxisymmetric sky distribution, we construct a toy hotspot
template,
\begin{equation}
\widetilde{P}_{\mathrm{toy}}(e,\lambda)=P_{0}+\sum_{k=1}^{N_{\mathrm{spot}}%
}A_{k}\exp\!\left[  -\frac{\zeta_{k}(e,\lambda)^{2}}{2\sigma_{k}^{2}}\right]
,\label{eq:toy_template}%
\end{equation}
where $P_{0}$ is a uniform background floor, $A_{k}$ is the amplitude of the
$k$th hotspot, and $\sigma_{k}$ characterizes its angular width. The quantity
$\zeta_{k}$ denotes the angular distance between the sky direction
$(e,\lambda)$ and the center $(e_{k},\lambda_{k})$ of the $k$th hotspot,
namely $\cos\zeta_{k}=\cos e\cos e_{k}+\sin e\sin e_{k}\cos(\lambda
-\lambda_{k})$. Thus, $\zeta_{k}$ measures how far a given sky position is
from a hotspot center, while the Gaussian factor in Eq.~(\ref{eq:toy_template}%
) ensures that each hotspot is localized on the sphere. We set $P_{0}=0.08$
and use the three tuples $(e_{k},\lambda_{k},\sigma_{k},A_{k})$
$=(1.00,0.70,0.28,3.0)$, $(2.10,3.90,0.22,2.2)$, and $(1.75,5.40,0.35,1.4)$,
with all angles in radians. As in the preceding sections, we divide
$\widetilde{P}_{\mathrm{toy}}$ by its full-sky integral to obtain
$P_{\mathrm{toy}}$, so that $\int_{S^{2}}d\Omega\,P_{\mathrm{toy}}(\Omega)=1$.
This toy model is not intended to represent a specific astrophysical source
population. Rather, it is introduced as a methodological benchmark with a much
stronger nonaxisymmetric component than the B20 Galactic template discussed
above. Its role is to test how the lunar response behaves when the sky
distribution contains several localized excesses distributed over different
directions, so that the projected quadrupolar covariance is expected to depart
more substantially from a nearly diagonal form.

\begin{figure}[t]
\centering
\includegraphics[width=0.95\linewidth]{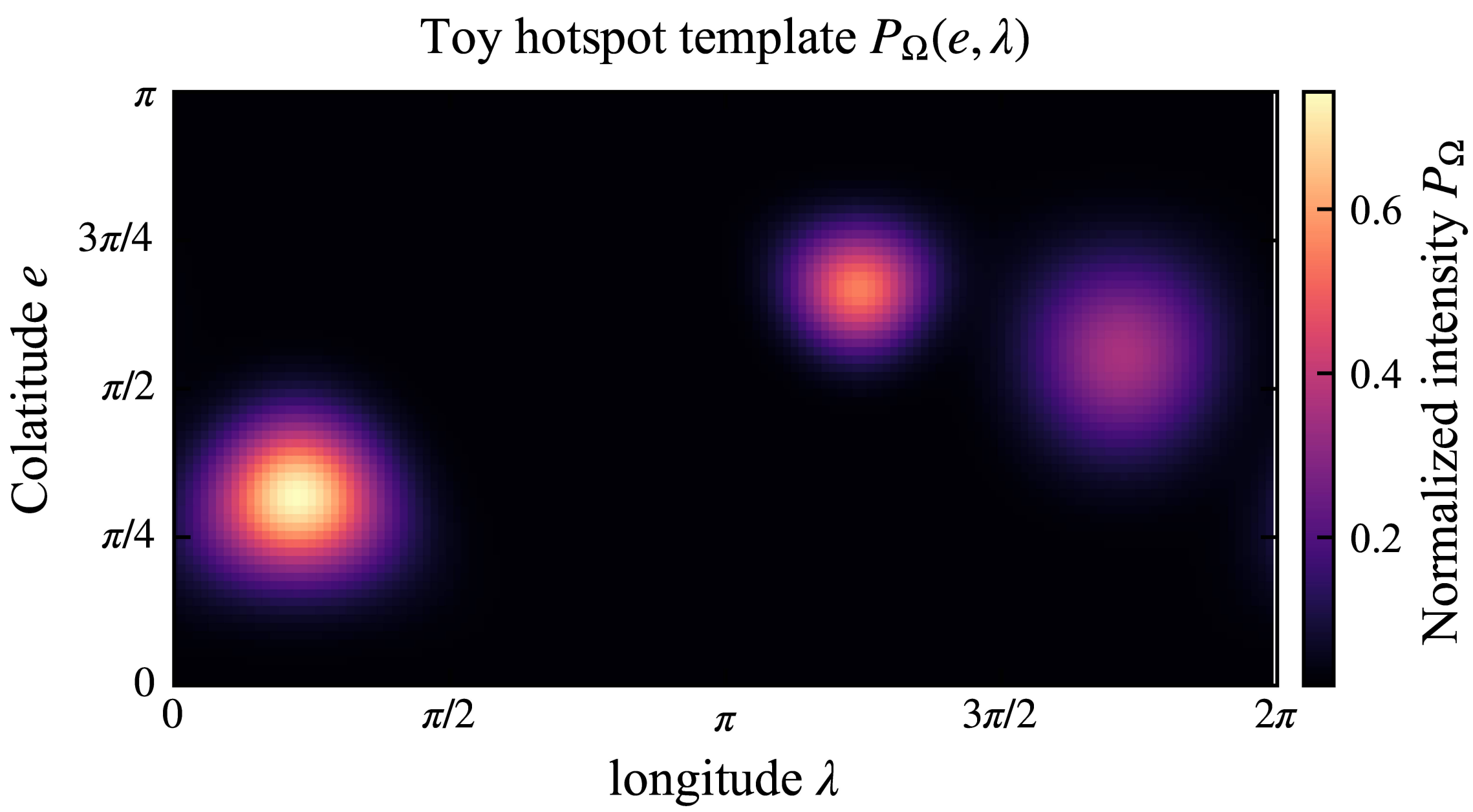}\caption{Toy hotspot sky
template $P(e,\lambda)$. The color-bar label $P_{\Omega}$ denotes the same
normalized sky intensity $P_{\mathrm{toy}}$. The map contains three localized
hotspots with different amplitudes and angular widths, superposed on a weak
uniform background.}%
\label{Fig9}%
\end{figure}

The corresponding template map is shown in Fig.~\ref{Fig9}. It is far from
uniform and contains three well-localized hotspots at different longitudes and
colatitudes. One hotspot is dominant, while the other two have lower
amplitudes and different angular widths. This means that the incident
stochastic background has clear preferred directions on the sky and is
therefore strongly anisotropic. It provides a controlled test of how
localized, nonaxisymmetric angular structure is projected into the lunar
quadrupolar response.

\begin{figure}[ptb]
\centering
\includegraphics[width=0.58\linewidth]{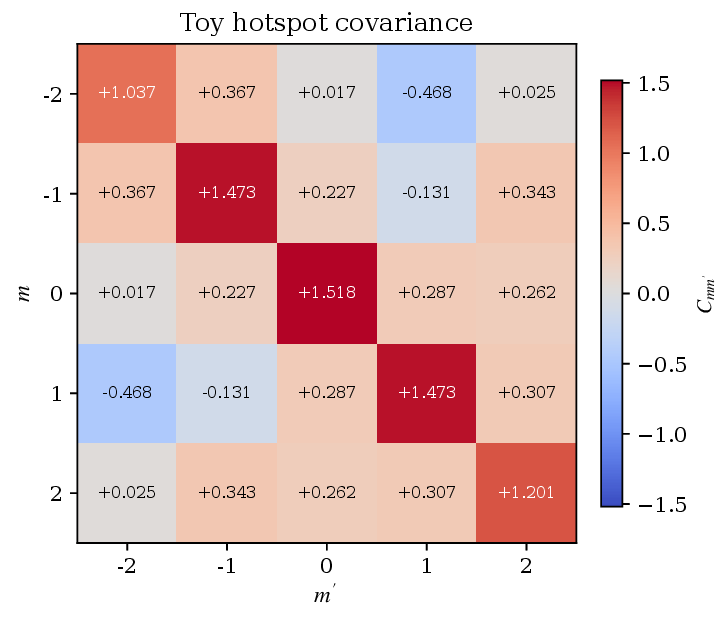}\caption{Projected covariance
matrix $C_{mm^{\prime}}$ for the toy hotspot template. The sizable positive
and negative off-diagonal components indicate substantially stronger
nonaxisymmetric quadrupolar coupling than in the B20 cases.}%
\label{Fig10}%
\end{figure}

For the toy hotspot template, the numerical evaluation yields the projected
covariance matrix shown in Fig.~\ref{Fig10}. The template satisfies
$\int_{S^{2}}P_{\mathrm{toy}}(\Omega)\,d\Omega=1$ to numerical precision. The
covariance matrix is clearly not diagonal: several off-diagonal components are
sizable, such as $C_{-2,-1}\simeq0.367$, $C_{-2,1}\simeq-0.468$,
$C_{-1,2}\simeq0.343$, and $C_{1,2}\simeq0.307$. Its diagonal entries also
differ, ranging from approximately $1.037$ to $1.518$. Unequal diagonal
entries are allowed for an anisotropic sky; it is the substantial off-diagonal
covariance, rather than diagonal inequality alone, that demonstrates mixing
between distinct lunar-frame quadrupolar channels.

Figure~\ref{Fig10} makes this strongly nondiagonal structure immediately
visible. The off-diagonal channel-mixing fraction, $\eta_{\mathrm{aniso}%
}\simeq0.383$, shows that the off-diagonal part of the projected covariance is
a substantial fraction of its total norm in the chosen lunar basis. This
matrix also shows why the lunar surface response does not reproduce the sky
map point by point. The template is first projected into the five-dimensional
$l=2$ covariance matrix and is then combined with the lunar kernels
$\boldsymbol{\mathcal{G}}_{2m}(\theta,\varphi;f)$ to produce the surface response.

Under a rotation of coordinates, the sky template, the covariance matrix and
the response basis transform together, and the scalar surface power is
unchanged. Consequently, a rotation-consistency check tests the numerical
implementation, whereas the nonzero off-diagonal covariance in
Fig.~\ref{Fig10} and the value of $\eta_{\mathrm{aniso}}$ quantify the
nonaxisymmetric mixing produced by the hotspot template. The isotropic null
test in Fig.~\ref{FigIso} provides the complementary numerical check.

\begin{figure}[t]
\centering
\includegraphics[width=0.96\textwidth]{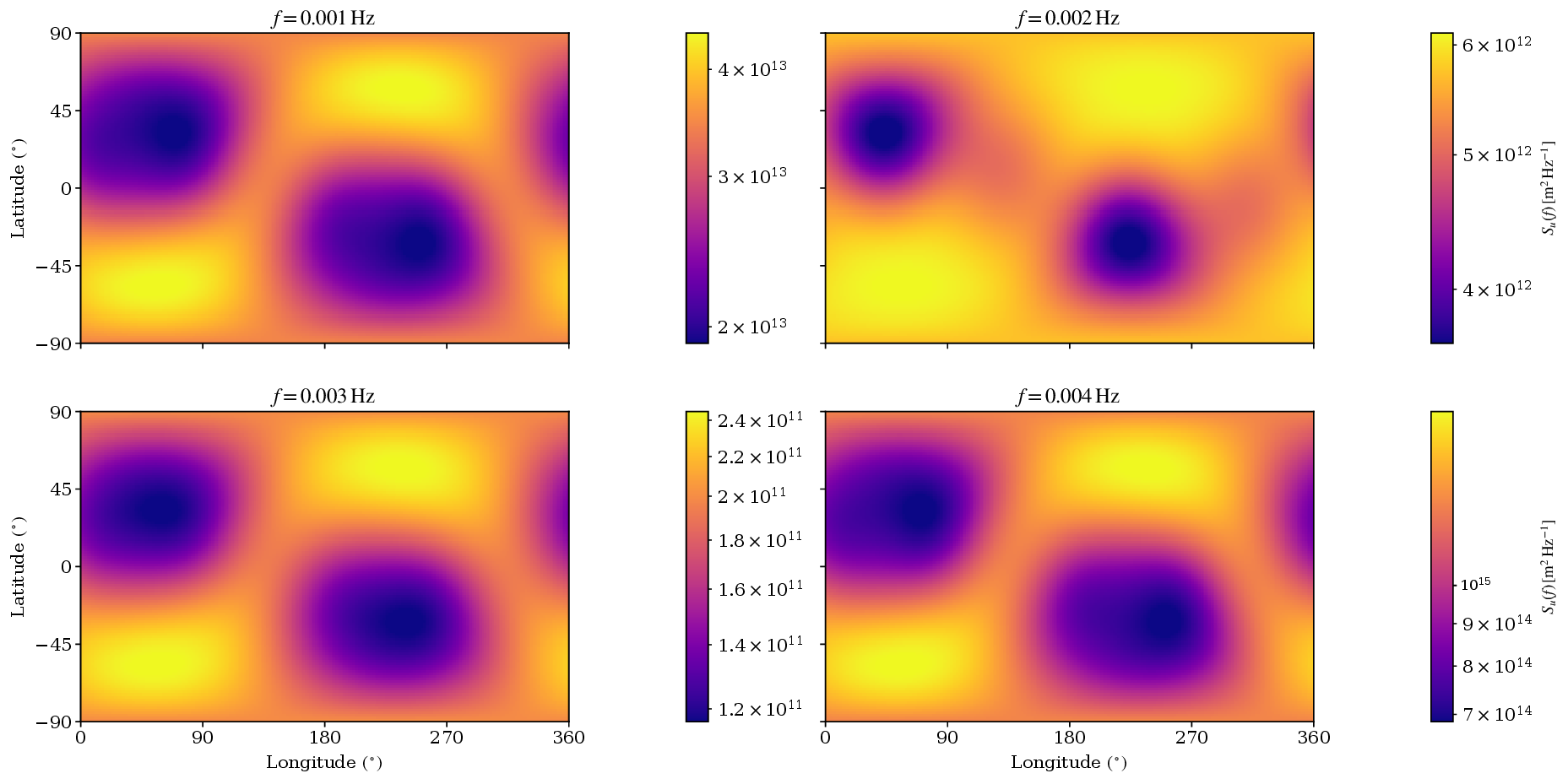}\caption{Hotspot-weighted
mean-square displacement spectral density $S_{u}(\theta,\varphi;f)$ at
$f=0.001$, $0.002$, $0.003$, and $0.004$~Hz for $S_{h}(f)=1\,\mathrm{Hz}^{-1}%
$. The distinct $0.002$~Hz morphology is associated with the enhanced
horizontal contribution near the $n=1$ mode at $f_{n}=1.917866$~mHz. The
color scales are in $\mathrm{m}^{2}\,\mathrm{Hz}^{-1}$.}%
\label{Fig11}%
\end{figure}

\begin{figure}[t]
\centering
\includegraphics[width=0.62\linewidth]{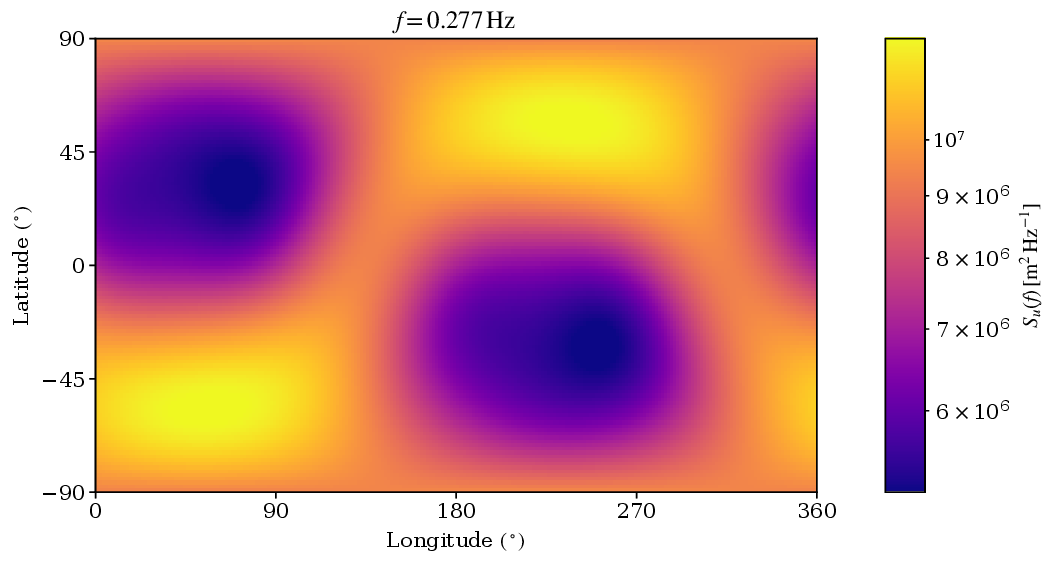}\caption{Hotspot-weighted
mean-square displacement spectral density $S_{u}(\theta,\varphi;f)$ at
$f=0.277$~Hz for $S_{h}(f)=1\,\mathrm{Hz}^{-1}$. This frequency exceeds the
upper eigenfrequency of the retained $n\leq400$ normal-mode spectrum; the map
is sensitive to normal-mode truncation. The color scale is in $\mathrm{m}%
^{2}\,\mathrm{Hz}^{-1}$.}%
\label{Fig12}%
\end{figure}

Figure~\ref{Fig11} shows the hotspot-weighted mean-square displacement
spectral-density maps at $f=0.001$, $0.002$, $0.003$, and $0.004$~Hz. The sky
template and its covariance matrix are fixed in all four panels; the frequency
dependence enters only through $\boldsymbol{\mathcal{G}}_{2m}(\theta
,\varphi;f)$. The sizable off-diagonal entries of $C_{mm^{\prime}}$ produce
two-dimensional surface structure, while the radial and tangential normal-mode
transfer functions determine how it appears at each frequency. In particular,
the distinct morphology near $0.002$~Hz is associated with the enhanced
horizontal response close to the $n=1$ mode. The map at $f=0.277$~Hz in
Fig.~\ref{Fig12} lies above the highest retained eigenfrequency,
$f_{400}=0.273038$~Hz. As for the B20 template, this map depends on the
normal-mode truncation (Sec.~\ref{sec:cutoff_sensitivity}).

Overall, the toy hotspot model provides a useful complement to the B20
Galactic template. The aligned B20 case has weak off-diagonal mixing among its
quadrupolar channels. The toy template demonstrates that a stronger
nonaxisymmetric sky component can produce substantial quadrupolar-channel
mixing and a two-dimensional lunar response. For a fixed $P(\Omega)$, its
quadrupolar projection is described by the frequency-independent covariance
$C_{mm^{\prime}}$. The frequency dependence of the surface response enters
through the lunar normal-mode kernels $\boldsymbol{\mathcal{G}}_{2m}%
(\theta,\varphi;f)$.

\section{Conclusion}

In this paper, we developed a framework for describing the response of the
Moon to an anisotropic SGWB using its quadrupolar normal-mode channels. The
angular distribution of the background is represented by a sky template
$P(\Omega)$, whose effect on the lunar response is summarized by the projected
covariance matrix $C_{mm^{\prime}}$ in the $l=2$ subspace. This matrix
connects the sky anisotropy to the five azimuthal quadrupolar response
channels of the Moon. Within this formulation, the lunar surface response is
not a direct geometrical image of the sky map; rather, it is determined by the
quadrupolar projection of the sky distribution together with the
frequency-dependent lunar normal-mode response kernels.

We applied this framework to two representative templates. For the
B20-inspired Galactic template, which represents the angular structure of an
unresolved Galactic foreground, the projected covariance matrix remains close
to diagonal in the aligned case. Thus, although the B20 sky map is visibly
anisotropic, its off-diagonal channel mixing is weak in the aligned lunar
frame. Its unequal diagonal entries still distinguish the projected B20
covariance from an isotropic sky, so the off-diagonal ratio alone is not a
complete measure of anisotropy. When the B20 foreground axis is tilted
relative to the lunar reference axis, the off-diagonal components of
$C_{mm^{\prime}}$ increase and the channel-mixing fraction $\eta
_{\mathrm{aniso}}$ becomes larger. This demonstrates that the projected lunar
response is sensitive not only to the angular structure of the sky template,
but also to its orientation relative to the lunar quadrupolar basis. The toy
hotspot template provides a complementary test with stronger nonaxisymmetric
structure. In this case, the projected covariance matrix develops substantial
off-diagonal components, and the corresponding $\eta_{\mathrm{aniso}}$ is much
larger than in the aligned B20 case. The resulting response maps show clear
two-dimensional, nonaxisymmetric surface patterns. This comparison shows that
the projected covariance can differ from the isotropic control even when
distinct-channel mixing is small, whereas localized, nonaxisymmetric angular
structure can produce appreciable mixing among the lunar quadrupolar response channels.

For the hotspot template, the surface morphology varies with frequency through
the radial and horizontal normal-mode transfer functions in the lunar response
kernels, while the sky template is fixed. In particular, the distinct hotspot
morphology near $f=0.002~\mathrm{Hz}$ in Fig.~\ref{Fig11} is associated with a
change in the relative weight of the radial and horizontal surface terms, with
the horizontal response enhanced near the $n=1$ mode at $f_{n}%
=1.917866~\mathrm{mHz}$. At $f=0.277~\mathrm{Hz}$, above the highest retained
eigenfrequency, the aligned-B20 surface spectral density changes strongly
between the $n\leq200$ and $n\leq400$ sums. The displayed map is therefore a
finite-mode illustration, not a converged high-frequency prediction.

The isotropic control recovers quadrupolar covariance diagonals equal within
angular-quadrature error and a uniform surface power map. The aligned B20
covariance has unequal diagonal entries, while tilting the same sky increases
its off-diagonal entries in the lunar basis. Together with the hotspot
comparison, these controls support the interpretation of the B20 diagonal
spread and hotspot channel mixing as template-driven, rather than artifacts of
the tested angular calculation.

Overall, the projected covariance matrix provides a compact way to connect
anisotropic SGWB sky templates with lunar surface-response morphologies. The
results show that the Moon can act as a directionally sensitive elastic
response system to anisotropic stochastic backgrounds, while also emphasizing
that quantitative conclusions at higher frequencies require convergence tests
with extended normal-mode sums.

\section{Acknowledgments}

We thank the anonymous referee for constructive comments that significantly
improved this paper. This work is supported by National Key Research and
Development Program of China with Grant No. 2024YFC2207300, National Natural
Science Foundation of China (NSFC) with Grant No. 12375057, and the
Fundamental Research Funds for the Central Universities, China University of
Geosciences (Wuhan).

\appendix

\section{DETAILED GW-ELASTIC RESPONSE FORMALISM}

\label{app:response-formalism}

This appendix derives the single-wave normal-mode response in Eq.~(\ref{rf2})
in four steps. Appendix~A\thinspace1 constructs the selenocentric frame and GW
polarization tensor; Appendix~A\thinspace2 introduces the Green tensor and
normal modes of a radially stratified sphere; Appendix~A\thinspace3 combines
these two ingredients in the long-wavelength limit to select $l=2$ and obtain
the five angular factors $F^{m}$; Appendix~A\thinspace4 introduces the damped
transfer function and recovers the real displacement.
Appendix~\ref{app:stochastic-response} then derives the lunar response
directly from the stochastic field of Eq.~(\ref{eq:sgwb_superposition}),
including the forcing, normal-mode projection, and reconstruction of the
physical displacement. It also recovers the deterministic single-wave response
as a special case. The coordinate and normal-mode constructions are
independent until they meet in Appendix~A\thinspace3. We follow the standard
treatments in global seismology and GW-elastic-body coupling
\cite{FJD1969,ABM1983,DT2020,MRR2019}.

\subsection{Selenocentric coordinates and polarization tensor}

\label{app:coordinates}

This subsection fixes the incident-direction and polarization conventions used
to construct the tensor $\boldsymbol{\epsilon}$. We adopt the IAU
selenocentric cartographic convention \cite{DAC1980,AAC2011}. The origin is at
the lunar center, $\hat{\boldsymbol{e}}_{x}$ points toward the mean Earth
direction, $\hat{\boldsymbol{e}}_{z}$ points toward the lunar north pole, and
$\hat{\boldsymbol{e}}_{y}$ completes a right-handed Cartesian triad. This
frame is illustrated in Fig.~\ref{Fig-a1}.

\renewcommand{\thefigure}{A\arabic{figure}} \setcounter{figure}{0}
\begin{figure}[ptb]
\centering
\includegraphics[width=0.32\linewidth]{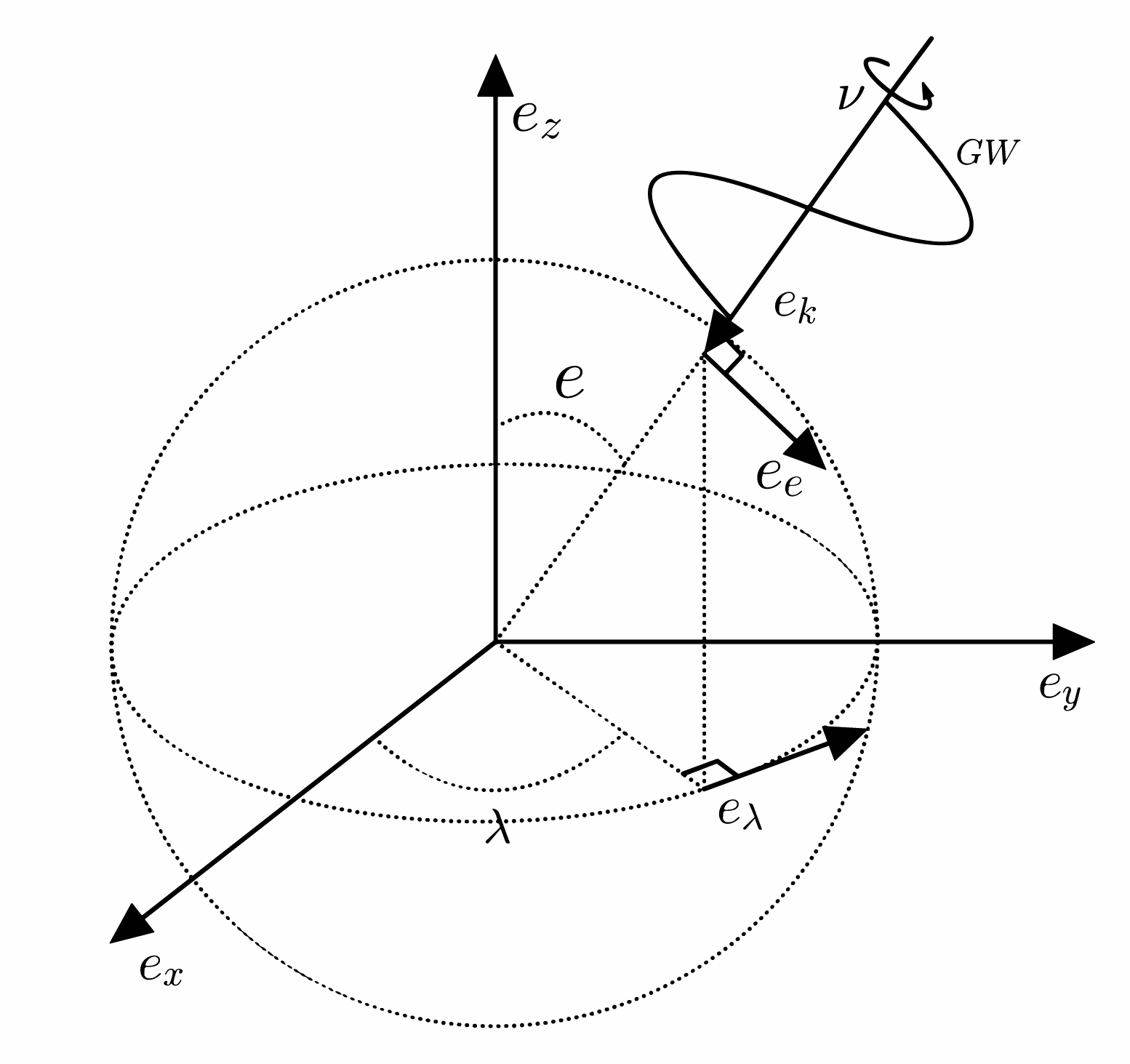}\caption{IAU selenocentric
coordinate system. The origin is at the lunar center; $\hat{\boldsymbol{e}%
}_{x}$ points toward the mean Earth direction, $\hat{\boldsymbol{e}}_{z}$
points toward the lunar north pole, and $\hat{\boldsymbol{e}}_{y}$ completes
the right-handed triad.}%
\label{Fig-a1}%
\end{figure}

The incident GW direction is parametrized by the colatitude $e\in\lbrack
0,\pi]$ and longitude $\lambda\in\lbrack0,2\pi)$,
\begin{equation}
\hat{\boldsymbol{k}}=(\sin e\cos\lambda,\,\sin e\sin\lambda,\,\cos e),
\end{equation}
where $e$ is measured from $+\hat{\boldsymbol{e}}_{z}$ and $\lambda$ from
$+\hat{\boldsymbol{e}}_{x}$ toward $+\hat{\boldsymbol{e}}_{y}$. The associated
orthonormal tangent vectors are
\begin{align}
\hat{\boldsymbol{e}}_{e} &  =(\cos e\cos\lambda,\,\cos e\sin\lambda,\,-\sin
e),\nonumber\\
\hat{\boldsymbol{e}}_{\lambda} &  =(-\sin\lambda,\,\cos\lambda,\,0).
\end{align}
Introducing the polarization angle $\nu$, two mutually orthogonal unit vectors
in the plane transverse to $\hat{\boldsymbol{k}}$ are
\begin{align}
\boldsymbol{I} &  =-\cos\nu\,\hat{\boldsymbol{e}}_{e}-\sin\nu\,\hat
{\boldsymbol{e}}_{\lambda},\nonumber\\
\boldsymbol{m} &  =\sin\nu\,\hat{\boldsymbol{e}}_{e}-\cos\nu\,\hat
{\boldsymbol{e}}_{\lambda}.
\end{align}
They satisfy $\boldsymbol{I}\cdot\boldsymbol{m}=0$ and $\boldsymbol{I}%
\cdot\hat{\boldsymbol{k}}=\boldsymbol{m}\cdot\hat{\boldsymbol{k}}=0$.

The plus and cross polarization tensors are \cite{MRR2019}
\begin{align}
\boldsymbol{e}_{+}  &  =\boldsymbol{I}\otimes\boldsymbol{I }-\boldsymbol{m}%
\otimes\boldsymbol{m},\nonumber\\
\boldsymbol{e}_{\times}  &  =\boldsymbol{I}\otimes\boldsymbol{m }%
+\boldsymbol{m}\otimes\boldsymbol{I}.
\end{align}
For the equal-power, unpolarized convention used to evaluate the angular
response factors in this work, we take the unit-amplitude representative
\begin{equation}
\boldsymbol{\epsilon}=\boldsymbol{e}_{+}+\boldsymbol{e}_{\times}.
\label{eq:epsilon_convention}%
\end{equation}
The average over $\nu\in[0,\pi]$ of this representative gives the same
equal-power, vanishing-cross-correlation weighting of the two polarizations as
Eq.~(\ref{anisotropic_psd}).

Writing $\boldsymbol{\epsilon}$ in the selenocentric Cartesian basis gives
\begin{equation}
\boldsymbol{\epsilon}=
\begin{pmatrix}
\epsilon_{xx} & \epsilon_{xy} & \epsilon_{xz}\\
\epsilon_{xy} & \epsilon_{yy} & \epsilon_{yz}\\
\epsilon_{xz} & \epsilon_{yz} & \epsilon_{zz}%
\end{pmatrix}
.
\end{equation}
With $\mathcal{A}_{\nu}=\cos2\nu-\sin2\nu$ and $\mathcal{B}_{\nu}=\cos
2\nu+\sin2\nu$, its independent components are
\begin{align}
\epsilon_{xx}  &  =\mathcal{A}_{\nu}(\cos^{2}e\cos^{2}\lambda-\sin^{2}\lambda)
-\mathcal{B}_{\nu}\cos e\sin2\lambda,\nonumber\\
\epsilon_{xy}  &  =\tfrac12\mathcal{A}_{\nu}\sin2\lambda(\cos^{2}e+1)
+\mathcal{B}_{\nu}\cos e\cos2\lambda,\nonumber\\
\epsilon_{xz}  &  =-\tfrac12\mathcal{A}_{\nu}\sin2e\cos\lambda+\mathcal{B}%
_{\nu}\sin e\sin\lambda,\nonumber\\
\epsilon_{yy}  &  =\mathcal{A}_{\nu}(\cos^{2}e\sin^{2}\lambda-\cos^{2}\lambda)
+\mathcal{B}_{\nu}\cos e\sin2\lambda,\nonumber\\
\epsilon_{yz}  &  =-\tfrac12\mathcal{A}_{\nu}\sin2e\sin\lambda-\mathcal{B}%
_{\nu}\sin e\cos\lambda,\nonumber\\
\epsilon_{zz}  &  =\mathcal{A}_{\nu}\sin^{2}e. \label{eq:eps_full}%
\end{align}
The remaining components follow from symmetry. Direct substitution verifies
$\epsilon_{ij}=\epsilon_{ji}$, $\epsilon_{ii}=0$, and $\epsilon_{ij}\hat
k_{j}=0$. Appendix~A\,3 projects these components onto the five-dimensional
quadrupolar subspace to obtain the angular coefficients $F^{m}(e,\lambda,\nu)$
used in the main text.

\subsection{Green tensor and normal modes}

This subsection supplies the normal-mode basis and Green tensor for the
radially stratified, nonrotating spherical model used in Sec.~II. For such an
elastic sphere, the displacement generated by the equivalent body force
$\boldsymbol{f}$ and surface traction $\boldsymbol{\tau}$ is
\begin{align}
\boldsymbol{u}(\boldsymbol{r},t) &  =\int_{-\infty}^{t}\!\!\int_{V}%
G(\boldsymbol{r},\boldsymbol{r}^{\prime};t-t^{\prime})\boldsymbol{f}%
(\boldsymbol{r}^{\prime},t^{\prime})\,dV^{\prime}\,dt^{\prime}\nonumber\\
&  \quad+\int_{-\infty}^{t}\!\!\int_{S}G(\boldsymbol{r},\boldsymbol{r}%
^{\prime};t-t^{\prime})\boldsymbol{\tau}(\boldsymbol{r}^{\prime},t^{\prime
})\,d\Sigma^{\prime}\,dt^{\prime},
\end{align}
where $G$ is the second-order Green tensor and $\boldsymbol{\tau}=\mu
(R)\hat{\boldsymbol{e}}_{r}\cdot\boldsymbol{h}$ enforces the free-surface
condition. The eigenfunctions solve the homogeneous equations of motion; weak
damping is included through diagonal normal-mode damping terms. With the
normal-mode mass norm $N_{k}$, the retarded Green tensor is
\begin{equation}
\begin{aligned} G(\boldsymbol r,\boldsymbol r';t) &=\sum_k\frac{\boldsymbol s_k(\boldsymbol r)\otimes \boldsymbol s_k^*(\boldsymbol r')}{N_k}\,g_k(t),\\ g_k(t)&=\Theta(t)e^{-\Gamma_k t} \frac{\sin(\omega_{d,k}t)}{\omega_{d,k}}. \end{aligned}\label{eq:retarded_green_tensor}%
\end{equation}
Here $\Gamma_{k}=\omega_{k}/(2Q_{k})$, $\omega_{d,k}=\sqrt{\omega_{k}%
^{2}-\Gamma_{k}^{2}}$ in the underdamped regime, $\Theta(t)$ is the Heaviside
step function, and $\otimes$ denotes the outer product. The scalar impulse
response $g_{k}$ vanishes for $t<0$, has $g_{k}(0^{+})=0$ and $\dot{g}%
_{k}(0^{+})=1$, and obeys $\ddot{g}_{k}+2\Gamma_{k}\dot{g}_{k}+\omega_{k}%
^{2}g_{k}=0$ for $t>0$. These conditions specify the causal Green function of
the damped normal-mode equation used in Appendix~\ref{app:stochastic-response}%
. The index $k=(n,l,m)$ labels radial order, angular degree, and azimuthal
order. The displacement eigenfunction is expanded in vector spherical
harmonics as
\begin{equation}
\boldsymbol{s}_{k}=U_{k}(r)\boldsymbol{P}_{ml}+V_{k}(r)\boldsymbol{B}%
_{ml}+W_{k}(r)\boldsymbol{C}_{ml},
\end{equation}
with
\begin{align}
\boldsymbol{P}_{ml} &  =\boldsymbol{e}_{r}Y_{lm},\nonumber\\
\sqrt{l(l+1)}\boldsymbol{B}_{ml} &  =\left(  \boldsymbol{e}_{\theta}%
\partial_{\theta}+\boldsymbol{e}_{\varphi}\frac{1}{\sin\theta}\partial
_{\varphi}\right)  Y_{lm},\nonumber\\
\sqrt{l(l+1)}\boldsymbol{C}_{ml} &  =\left(  \boldsymbol{e}_{\theta}\frac
{1}{\sin\theta}\partial_{\varphi}-\boldsymbol{e}_{\varphi}\partial_{\theta
}\right)  Y_{lm}.
\end{align}
Here $U_{k}$, $V_{k}$, and $W_{k}$ are the radial, spheroidal tangential, and
toroidal eigenfunctions. We restrict the calculation to spheroidal modes and
set $W_{k}=0$. With the normalized angular basis above, their mass norm is
\begin{equation}
N_{k}=\int_{0}^{R}\rho(r)r^{2}[U_{k}^{2}(r)+V_{k}^{2}(r)]\,dr.
\end{equation}
At the surface $r=R$, the quadrupolar spheroidal eigenfunction is
\begin{align}
\boldsymbol{s}_{n2m}(R,\theta,\varphi) &  =U_{n}(R)Y_{2m}\hat{\boldsymbol{e}%
}_{r}+\frac{V_{n}(R)}{\sqrt{6}}\partial_{\theta}Y_{2m}\hat{\boldsymbol{e}%
}_{\theta}\nonumber\\
&  \quad+\frac{V_{n}(R)}{\sqrt{6}\sin\theta}\partial_{\varphi}Y_{2m}%
\hat{\boldsymbol{e}}_{\varphi}.
\end{align}

Here $\boldsymbol{s}_{n2m}(R,\theta,\varphi)$ is the eigenfunction evaluated
at the lunar surface. The angles $(\theta,\varphi)$ are the colatitude and
longitude of the observation point, whereas $\Omega=(e,\lambda)$ specifies the
incident GW direction in Appendix~A\,1. The lunar spin axis is the reference
$z$-axis in our maps; the rotation in Sec.~V permits another preferred axis.

Before applying the time-dependent transfer function, we project the
unit-strain volume force and free-surface traction onto a mode $k$,
\begin{align}
\mathcal{C}_{k} &  =-\int_{V}\frac{d\mu}{dr}\boldsymbol{s}_{k}^{\ast
}(\boldsymbol{r}^{\prime})\cdot\left(  \boldsymbol{\epsilon}\cdot
\hat{\boldsymbol{e}}_{r}\,e^{-i\boldsymbol{k}_{g}\cdot\boldsymbol{r}^{\prime}%
}\right)  dV^{\prime}\nonumber\\
&  \quad+\mu(R)\int_{S}\boldsymbol{s}_{k}^{\ast}(\boldsymbol{r}^{\prime}%
)\cdot\left(  \boldsymbol{\epsilon}\cdot\hat{\boldsymbol{e}}_{r}%
\,e^{-i\boldsymbol{k}_{g}\cdot\boldsymbol{r}^{\prime}}\right)  d\Sigma
^{\prime}.\label{sk_align}%
\end{align}
Here $R$ is the outer radius. The factor $h_{0}e^{i\omega_{g}t}$ has been
separated from this source overlap; it is restored in Appendix~A\thinspace4.
The lunar case uses the lunar profiles $\rho(r)$ and $\mu(r)$.

\subsection{Angular reduction and radial coupling}

\label{app:angular-reduction}

This subsection combines the source projection of Eq.~(\ref{sk_align}) with
the polarization tensor of Appendix~A\,1. We assume the long-wavelength
condition $k_{g}R\ll1$, so the phase factor is approximately unity throughout
the sphere. We use $d\Omega_{r}=\sin\theta\,d\theta\,d\varphi$ for the
solid-angle element of a lunar radial direction and $\nabla_{S}$ for the
gradient on that unit sphere; $\Omega=(e,\lambda)$ continues to denote the
incident GW direction. Let $\hat{\boldsymbol{r}}=(x,y,z)$ be a unit radial
vector and set $q(\hat{\boldsymbol{r}})=\epsilon_{ij}\hat r_{i}\hat r_{j}$.
Because $\epsilon_{ii}=0$, this is the restriction of a harmonic quadratic
Cartesian polynomial to the unit sphere; it therefore has only $l=2$ angular
content. The vector entering Eq.~(\ref{sk_align}) separates into radial and
tangential parts as
\begin{equation}
\boldsymbol{\epsilon}\cdot\hat{\boldsymbol{r}} =q\hat{\boldsymbol{r}}%
+\tfrac12\nabla_{S} q, \qquad\nabla_{S} q =2(\boldsymbol{\epsilon}\cdot
\hat{\boldsymbol{r}} -q\hat{\boldsymbol{r}}).
\end{equation}
Thus both parts of the spheroidal-mode overlap are determined by the same
quadrupolar scalar $q$.

We use the orthonormal real $l=2$ basis paired with the angular coefficients
in the main text:
\begin{align}
Y_{20}  &  =\sqrt{\frac{5}{16\pi}}(3z^{2}-1), & Y_{2,-2}  &  =\sqrt{\frac
{15}{16\pi}}(x^{2}-y^{2}),\nonumber\\
Y_{2,-1}  &  =-\sqrt{\frac{15}{4\pi}}xz, & Y_{2,1}  &  =-\sqrt{\frac{15}{4\pi
}}yz,\nonumber\\
Y_{2,2}  &  =\sqrt{\frac{15}{4\pi}}xy. &
\end{align}
Each basis function is quadratic in $(x,y,z)$, and the five functions span the
symmetric trace-free quadrupolar subspace. Defining
\begin{equation}
\Lambda_{ij}^{(m)}=\int_{S^{2}}d\Omega_{r}\,Y_{2m}(\hat{\boldsymbol{r}}) \hat
r_{i}\hat r_{j},
\end{equation}
gives the five real symmetric trace-free tensors used in Eq.~(\ref{rf2}). For
example, the $Y_{20}$ projection gives $\Lambda_{zz}^{(0)}=(4/3)\sqrt{\pi/5}$
and $\Lambda_{xx}^{(0)}=\Lambda_{yy}^{(0)}=-(2/3)\sqrt{\pi/5}$; the remaining
entries follow from the same integral. In matrix form they are
\begin{align}
\Lambda^{(0)}  &  = \frac{2}{3}\sqrt{\frac{\pi}{5}}
\begin{pmatrix}
-1 & 0 & 0\\
0 & -1 & 0\\
0 & 0 & 2
\end{pmatrix}
,\\
\Lambda^{(-2)}  &  =2\sqrt{\frac{\pi}{15}}
\begin{pmatrix}
1 & 0 & 0\\
0 & -1 & 0\\
0 & 0 & 0
\end{pmatrix}
,\qquad\Lambda^{(2)}=2\sqrt{\frac{\pi}{15}}
\begin{pmatrix}
0 & 1 & 0\\
1 & 0 & 0\\
0 & 0 & 0
\end{pmatrix}
,\\
\Lambda^{(-1)}  &  =2\sqrt{\frac{\pi}{15}}
\begin{pmatrix}
0 & 0 & -1\\
0 & 0 & 0\\
-1 & 0 & 0
\end{pmatrix}
,\qquad\Lambda^{(1)}=2\sqrt{\frac{\pi}{15}}
\begin{pmatrix}
0 & 0 & 0\\
0 & 0 & -1\\
0 & -1 & 0
\end{pmatrix}
.
\end{align}
The directional factors are the contractions
\begin{equation}
F^{m}(\Omega,\nu)=\epsilon_{ij}\Lambda_{ij}^{(m)}, \qquad m=0,\pm1,\pm2.
\end{equation}
Equivalently, orthonormality gives $q(\hat{\boldsymbol{r}})=\sum_{m=-2}%
^{2}F^{m}Y_{2m}(\hat{\boldsymbol{r}})$. For a spheroidal eigenfunction, the
radial angular integral is $\int d\Omega_{r}\,Y_{lm}^{*}q=\delta_{l2}F^{m}$.
The tangential integral follows by integration by parts on the sphere:
\begin{align}
\int d\Omega_{r}\,\nabla_{S} Y_{lm}^{*}\cdot(\boldsymbol{\epsilon}\cdot
\hat{\boldsymbol{r}})  &  =\frac12\int d\Omega_{r}\,\nabla_{S} Y_{lm}^{*}
\cdot\nabla_{S} q\nonumber\\
&  =\frac{l(l+1)}{2}\int d\Omega_{r}\,Y_{lm}^{*}q =3\delta_{l2}F^{m}.
\end{align}
The factor $V_{n}/\sqrt6$ in the tangential eigenfunction consequently
produces $3V_{n}/\sqrt6$ in the radial overlap. These two identities establish
the $l=2$ selection rule for the source projection. Applying these angular
identities to the volume and surface terms in Eq.~(\ref{sk_align}) gives
\begin{align}
\mathcal{C}_{nlm}  &  =-\int_{0}^{R}\frac{d\mu}{dr}r^{2}dr\int d\Omega_{r}\,
\boldsymbol{s}_{nlm}^{*}\cdot(\boldsymbol{\epsilon}\cdot\hat{\boldsymbol{r}%
})\nonumber\\
&  \quad+\mu(R)R^{2}\int d\Omega_{r}\, \boldsymbol{s}_{nlm}^{*}\cdot
(\boldsymbol{\epsilon}\cdot\hat{\boldsymbol{r}})\nonumber\\
&  =\delta_{l2}F^{m}(\Omega,\nu)\,\mathcal{R}_{n},
\end{align}
where
\begin{equation}
\mathcal{R}_{n}=\mu(R)R^{2} \left[  U_{n}(R)+\frac{3}{\sqrt6}V_{n}(R)\right]
-\int_{0}^{R}\frac{d\mu}{dr} \left[  U_{n}(r)+\frac{3}{\sqrt6}V_{n}(r)\right]
r^{2}dr. \label{eq:radial_forcing_overlap}%
\end{equation}
For the TT polarization convention used here, the five angular factors are
\begin{align}
F^{m}(\Omega,\nu)  &  =\delta_{m,-2}\,2\sqrt{\tfrac{\pi}{15}} [\mathcal{A}%
_{\nu}\cos2\lambda(\cos^{2}e+1) -2\mathcal{B}_{\nu}\cos e\sin2\lambda
]\nonumber\\
&  \quad-\delta_{m,-1}\,4\sqrt{\tfrac{\pi}{15}}\sin e [\mathcal{B}_{\nu}%
\sin\lambda-\mathcal{A}_{\nu}\cos e\cos\lambda]\nonumber\\
&  \quad+\delta_{m,0}\,2\sqrt{\tfrac{\pi}{5}} \mathcal{A}_{\nu}\sin
^{2}e\nonumber\\
&  \quad+\delta_{m,1}\,4\sqrt{\tfrac{\pi}{15}}\sin e [\mathcal{B}_{\nu}%
\cos\lambda+\mathcal{A}_{\nu}\cos e\sin\lambda]\nonumber\\
&  \quad+\delta_{m,2}\,\sqrt{\tfrac{\pi}{15}} [4\mathcal{B}_{\nu}\cos
e\cos2\lambda+\mathcal{A}_{\nu}(3+\cos2e)\sin2\lambda],
\end{align}
where $\mathcal{A}_{\nu}=\cos2\nu-\sin2\nu$ and $\mathcal{B}_{\nu}=\cos
2\nu+\sin2\nu$.

The overlap $\mathcal{R}_{n}$ is the forcing projected onto the unnormalized
eigenfunction. Dividing it by the mass norm defined in Appendix~A\,2 gives the
coefficient that multiplies the eigenfunction in the response:
\begin{equation}
\alpha_{n}=\frac{\mathcal{R}_{n}}{N_{n}},\qquad N_{n}=\int_{0}^{R}
\rho(r)r^{2}[U_{n}^{2}(r)+V_{n}^{2}(r)]\,dr. \label{rs}%
\end{equation}
This $\alpha_{n}$ enters the single-mode displacement below and the radial and
tangential transfer sums in Eqs.~(\ref{A}) and (\ref{B}). Since $\mathcal{R}%
_{n}$ is linear and $N_{n}$ is quadratic in the eigenfunction, $\alpha
_{n}\boldsymbol{s}_{n2m}$ does not depend on its arbitrary overall
normalization. This is the same convention as Eq.~(\ref{rf2}).

\subsection{Damped transfer functions and real displacement}

\label{app:transfer-functions}

This subsection combines the normalized source coupling $\alpha_{n}$ with the
damped normal-mode source function and takes the real part after constructing
the complete response. For a monochromatic GW proportional to $e^{i\omega
_{g}t}$ we use the standard weak-damping forced-oscillator function
\cite{DT2020,MRR2019},
\begin{equation}
D_{n}(\omega_{g})=\frac{1}{\omega_{n}^{2}-\omega_{g}^{2}+i\omega_{n}\omega
_{g}/Q_{n}}.\label{eq:normal_mode_transfer}%
\end{equation}
Here the damping is specified by the quality factor $Q_{n}$. In this
subsection the incident parameters $(f,\Omega,\nu)$ are fixed and suppressed
in the displacement arguments. A single quadrupolar mode therefore
contributes
\begin{equation}
\boldsymbol{u}_{n2m}(\theta,\varphi,t)=h_{0}\Re\left\{  e^{i\omega_{g}t}%
D_{n}(\omega_{g})\alpha_{n}F^{m}(\Omega,\nu)\boldsymbol{s}_{n2m}%
(R,\theta,\varphi)\right\}  .
\end{equation}

After the selection rule is applied, $k=(n,2,m)$. In a nonrotating spherical
model, $U_{n}$, $V_{n}$, $\omega_{n}$, $Q_{n}$, and $\alpha_{n}$ are
independent of $m$ for a given radial order. Summing the modes gives
\begin{align}
\boldsymbol{u}_{\mathrm{coh}}(\theta,\varphi,t) &  =h_{0}\Re\Bigg\{e^{i\omega
_{g}t}\sum_{m=-2}^{2}F^{m}(\Omega,\nu)\left[  A(\omega_{g})Y_{2m}%
\hat{\boldsymbol{e}}_{r}\right.  \nonumber\\
&  \quad+B(\omega_{g})\partial_{\theta}Y_{2m}\hat{\boldsymbol{e}}_{\theta
}\left.  +\frac{B(\omega_{g})}{\sin\theta}\partial_{\varphi}Y_{2m}%
\hat{\boldsymbol{e}}_{\varphi}\right]  \Bigg\},
\end{align}
where
\begin{align}
A(\omega_{g}) &  :=\sum_{n}U_{n}(R)\alpha_{n}\frac{1}{\omega_{n}^{2}%
-\omega_{g}^{2}+i\omega_{n}\omega_{g}/Q_{n}},\label{A}\\
B(\omega_{g}) &  :=\frac{1}{\sqrt{6}}\sum_{n}V_{n}(R)\alpha_{n}\frac{1}%
{\omega_{n}^{2}-\omega_{g}^{2}+i\omega_{n}\omega_{g}/Q_{n}}.\label{B}%
\end{align}
The physical displacement is obtained by taking the real part only after the
complete normal-mode sum has been formed.

\subsection{Response to a stochastic gravitational-wave field}

\label{app:stochastic-response}

We now derive Eq.~(\ref{eq:stochastic_response_amplitude}) directly from the
random metric perturbation in Eq.~(\ref{eq:sgwb_superposition}). The
calculation retains the random strain amplitudes throughout the forcing and
normal-mode projection. Statistical averaging is needed only when forming the
displacement covariance. We use the same linear, nonrotating, spherically
symmetric lunar model and spheroidal normal modes as above.

\subsubsection{Random field and driving forces}

With $\omega_{g}=2\pi f>0$, introduce the positive-frequency tensor amplitude
$\widetilde h_{ij}(\boldsymbol{r};f)$ by writing
Eq.~(\ref{eq:sgwb_superposition}) as
\begin{align}
h_{ij}(\boldsymbol{r},t)  &  =\sqrt2\,\Re\int_{0}^{\infty}df\, e^{i\omega_{g}
t}\widetilde h_{ij}(\boldsymbol{r};f),\nonumber\\
\widetilde h_{ij}(\boldsymbol{r};f)  &  =\sum_{A=+,\times}\int_{S^{2}}%
d\Omega\, \widetilde h_{A}(f,\Omega)e^{A}_{ij}(\Omega) e^{-i\boldsymbol{k}%
_{f}(\Omega)\cdot\boldsymbol{r}}. \label{eq:app_random_field}%
\end{align}
The same positive-frequency convention is used for the force, traction, and
displacement below. In particular, $\widetilde h_{A}$ remains a random
amplitude, rather than a power spectral density. The fixed polarization
tensors $e^{A}_{ij}(\Omega)$ depend on the propagation direction, while the
position $\boldsymbol{r}$ lies inside the Moon.

The TT-gauge volume force and surface traction are linear in the metric
perturbation. Their frequency components therefore satisfy
\begin{align}
\widetilde f_{i}(\boldsymbol{r};f)  &  =-\mu^{\prime}(r)\hat r_{j}\widetilde
h_{ij}(\boldsymbol{r};f),\nonumber\\
\widetilde\tau_{i}(R,\hat{\boldsymbol{r}};f)  &  =\mu(R)\hat r_{j}\widetilde
h_{ij}(R\hat{\boldsymbol{r}};f). \label{eq:app_random_forcing}%
\end{align}
Here $\mu^{\prime}(r)=d\mu/dr$, and $\mu(R)$ denotes the value just inside the
surface. The volume term includes interior radial variations and interface
jumps of the shear modulus; the outer-surface contribution is included
separately through the traction. Substituting Eq.~(\ref{eq:app_random_field})
into Eq.~(\ref{eq:app_random_forcing}) thus places each $\widetilde
h_{A}(f,\Omega)$ explicitly in both driving terms. No average over sky
direction or polarization is taken at this stage.

\subsubsection{Normal-mode projection}

Expand the displacement in the normal modes of Appendix~A\thinspace2,
\begin{align}
\boldsymbol{u}(\boldsymbol{r},t) &  =\sum_{k}q_{k}(t)\boldsymbol{s}%
_{k}(\boldsymbol{r}),\nonumber\\
\int_{V}dV\,\rho\,\boldsymbol{s}_{k}^{\ast}\cdot\boldsymbol{s}_{k^{\prime}} &
=N_{k}\delta_{kk^{\prime}},\qquad k=(n,l,m).\label{eq:app_mode_expansion}%
\end{align}
Projecting the elastic boundary-value problem onto $\boldsymbol{s}_{k}^{\ast}$
and integrating by parts gives a volume-force term and a surface-traction
term. Orthogonality separates the undamped normal modes. With weak, diagonal
normal-mode damping represented by the quality factor $Q_{k}$, the amplitude
equation is
\begin{align}
\ddot{q}_{k}+\frac{\omega_{k}}{Q_{k}}\dot{q}_{k}+\omega_{k}^{2}q_{k} &
=\frac{b_{k}(t)}{N_{k}},\nonumber\\
b_{k}(t) &  =\int_{V}dV\,\boldsymbol{s}_{k}^{\ast}\cdot\boldsymbol{f}+\int
_{S}d\Sigma\,\boldsymbol{s}_{k}^{\ast}\cdot\boldsymbol{\tau}%
.\label{eq:app_forced_mode}%
\end{align}
The eigenfunctions satisfy the homogeneous free-surface condition; the
GW-induced traction appears in $b_{k}$ as an external driving term. Thus,
$b_{k}(t)$ is the generalized force on mode $k$, while $g_{k}(t)$ in
Eq.~(\ref{eq:retarded_green_tensor}) describes its response to an impulse.
Substituting the Green-tensor expansion into the volume and surface integrals
in Appendix~A\thinspace2 gives $b_{k}$ from the spatial integrals and the
causal driven amplitude from the remaining time convolution:
\[
q_{k}(t)=\frac{1}{N_{k}}\int_{-\infty}^{t}dt^{\prime}\,g_{k}(t-t^{\prime
})b_{k}(t^{\prime}).
\]
The physical displacement follows by summing $q_{k}(t)\boldsymbol{s}%
_{k}(\boldsymbol{r})$ over the modes. No free homogeneous oscillation is added
to this driven solution.

Using Eqs.~(\ref{eq:app_random_field}) and (\ref{eq:app_random_forcing}), the
frequency component of $b_{k}$ becomes
\begin{equation}
\widetilde b_{k}(f)=\sum_{A}\int_{S^{2}}d\Omega\, \widetilde h_{A}%
(f,\Omega)\,\mathcal{C}_{kA}(\Omega,f), \label{eq:app_random_mode_force}%
\end{equation}
where the source overlap is
\begin{align}
\mathcal{C}_{kA}(\Omega,f) ={}  &  -\int_{V} dV\,\mu^{\prime}(r)\boldsymbol{s}%
_{k}^{*}\cdot\bigl[\boldsymbol{e}^{A}(\Omega)\cdot\hat{\boldsymbol{r}%
}\bigr] e^{-i\boldsymbol{k}_{f}\cdot\boldsymbol{r}}\nonumber\\
&  +\mu(R)\int_{S} d\Sigma\,\boldsymbol{s}_{k}^{*}\cdot\bigl[\boldsymbol{e}%
^{A}(\Omega)\cdot\hat{\boldsymbol{r}}\bigr] e^{-i\boldsymbol{k}_{f}%
\cdot\boldsymbol{r}}. \label{eq:app_polarized_overlap}%
\end{align}
The sky integral in Eq.~(\ref{eq:app_random_mode_force}) sums incident
directions, whereas the integrals in Eq.~(\ref{eq:app_polarized_overlap})
project the force over the lunar volume and surface. These are distinct integrations.

Writing $q_{k}(t)=\sqrt2\Re\int_{0}^{\infty}df\, e^{i\omega_{g} t}\widetilde
q_{k}(f)$ and using $\partial_{t}\to i\omega_{g}$ in
Eq.~(\ref{eq:app_forced_mode}) gives the steady forced solution
\begin{align}
\widetilde q_{k}(f)  &  =\frac{D_{k}(\omega_{g})}{N_{k}}\widetilde
b_{k}(f),\nonumber\\
D_{k}(\omega_{g})  &  =\int_{0}^{\infty}dt\,g_{k}(t)e^{-i\omega_{g}
t}\nonumber\\
&  =\frac{1}{\omega_{k}^{2}-\omega_{g}^{2}+i\omega_{g}\omega_{k}/Q_{k}}.
\label{eq:app_random_mode_transfer}%
\end{align}
Free transients are omitted. For $k=(n,2,m)$, this transfer function is the
$m$-independent $D_{n}$ of Eq.~(\ref{eq:normal_mode_transfer}). The positive
sign of the damping term follows from the $e^{i\omega_{g} t}$ convention. The
integral form follows from the time convolution above; using $\omega_{d,k}%
^{2}+\Gamma_{k}^{2}=\omega_{k}^{2}$ gives the same denominator as the
frequency-domain normal-mode equation.

\subsubsection{Quadrupolar reduction and surface displacement}

In the long-wavelength regime $k_{f}R\ll1$, the spatial phase in
Eq.~(\ref{eq:app_polarized_overlap}) can be replaced by unity. For each
polarization, the angular reduction in Appendix~\ref{app:angular-reduction}
applies with $\boldsymbol{\epsilon}$ replaced by $\boldsymbol{e}^{A}(\Omega)$.
As in Appendix~\ref{app:angular-reduction}, $d\Omega_{r}$ is the solid-angle
element of the lunar radial direction and $\nabla_{S}$ is the gradient on that
unit sphere. The incident direction is denoted by $\Omega$. Define
\begin{equation}
\chi_{A}(\hat{\boldsymbol{r}};\Omega) =e^{A}_{ij}(\Omega)\hat r_{i}\hat
r_{j},\qquad F_{A}^{m}(\Omega)=e^{A}_{ij}(\Omega)\Lambda_{ij}^{(m)}.
\label{eq:app_polarization_projection}%
\end{equation}
The trace-free tensor $e^{A}_{ij}$ makes $\chi_{A}$ a pure quadrupole, and
$\boldsymbol{e}^{A}\cdot\hat{\boldsymbol{r}} =\chi_{A}\hat{\boldsymbol{r}%
}+\tfrac12\nabla_{S}\chi_{A}$. Consequently the radial and tangential angular
integrals are
\begin{align}
\int d\Omega_{r}\,Y_{lm}^{*}\chi_{A}  &  =\delta_{l2}F_{A}^{m},\nonumber\\
\int d\Omega_{r}\,\nabla_{S}Y_{lm}^{*}\cdot(\boldsymbol{e}^{A}\cdot
\hat{\boldsymbol{r}})  &  =\frac12\int d\Omega_{r}\,\nabla_{S}Y_{lm}^{*}%
\cdot\nabla_{S}\chi_{A}\nonumber\\
&  =\frac{l(l+1)}2\delta_{l2}F_{A}^{m} =3\delta_{l2}F_{A}^{m}.
\label{eq:app_angular_integrals}%
\end{align}
The second identity follows by integration by parts on the sphere. Together
with the tangential eigenfunction factor $1/\sqrt{l(l+1)}$, these identities
give
\begin{equation}
\mathcal{C}_{nlm,A}(\Omega,f) \simeq\delta_{l2}\,\mathcal{R}_{n}\,F_{A}%
^{m}(\Omega). \label{eq:app_reduced_overlap}%
\end{equation}
The radial overlap $\mathcal{R}_{n}$ is given in
Eq.~(\ref{eq:radial_forcing_overlap}); it contains both the surface term and
the interior integral, with the combination $U_{n}+3V_{n}/\sqrt6$. Its
normalization is $\alpha_{n}=\mathcal{R}_{n}/N_{n}$, as in Eq.~(\ref{rs}). The
approximation in Eq.~(\ref{eq:app_reduced_overlap}) is the long-wavelength
approximation used in the remaining expressions. Before this approximation,
$\mathcal{C}_{kA}(\Omega,f)$ depends on frequency through the spatial phase.
Its leading long-wavelength expression is frequency independent; the response
retains its frequency dependence through $D_{n}(\omega_{g})$.

Substitution into Eqs.~(\ref{eq:app_random_mode_force}) and
(\ref{eq:app_random_mode_transfer}) yields
\begin{equation}
\widetilde q_{n2m}(f)=D_{n}(\omega_{g})\alpha_{n} \sum_{A}\int_{S^{2}}%
d\Omega\, \widetilde h_{A}(f,\Omega)F_{A}^{m}(\Omega).
\label{eq:app_random_quadrupole_amplitude}%
\end{equation}
The random strain coefficients have therefore passed from the metric
perturbation to the normal-mode amplitudes without statistical averaging. All
spheroidal modes with $l\neq2$ have zero overlap at this order.

At the lunar surface, collect the radial-order sum into
\begin{equation}
\boldsymbol{\mathcal{G}}_{2m}(\theta,\varphi;f) =\sum_{n}D_{n}(\omega
_{g})\alpha_{n} \boldsymbol{s}_{n2m}(R,\theta,\varphi).
\label{eq:app_kernel_mode_sum}%
\end{equation}
Inserting the surface eigenfunction from Appendix~A\,2 gives
Eq.~(\ref{eq:G2m_def}), with $A(\omega_{g})=\sum_{n}D_{n}(\omega_{g}%
)\alpha_{n}U_{n}(R)$ and $B(\omega_{g})=\sum_{n}D_{n}(\omega_{g})\alpha
_{n}V_{n}(R)/\sqrt6$, as in Eqs.~(\ref{A}) and (\ref{B}). Summing
Eq.~(\ref{eq:app_random_quadrupole_amplitude}) over the normal modes now
gives
\begin{align}
\widetilde{\boldsymbol{u}}(\theta,\varphi;f)  &  =\sum_{n,m}\widetilde
q_{n2m}(f) \boldsymbol{s}_{n2m}(R,\theta,\varphi)\nonumber\\
&  =\sum_{A}\int_{S^{2}}d\Omega\,\widetilde h_{A}(f,\Omega) \sum_{m=-2}%
^{2}\boldsymbol{\mathcal{G}}_{2m}(\theta,\varphi;f) F_{A}^{m}(\Omega).
\label{eq:app_stochastic_surface_amplitude}%
\end{align}
This is Eq.~(\ref{eq:stochastic_response_amplitude}). It is a frequency-domain
random amplitude, not an averaged displacement or a power spectral density.
The physical displacement is reconstructed as
\begin{align}
\boldsymbol{u}(\theta,\varphi,t)  &  =\sqrt2\Re\int_{0}^{\infty}df\,
e^{i\omega_{g} t}\widetilde{\boldsymbol{u}}(\theta,\varphi;f)\nonumber\\
&  =\sqrt2\Re\left\{  \sum_{A}\int_{0}^{\infty}df\int_{S^{2}}d\Omega\,
\widetilde h_{A}(f,\Omega)e^{i\omega_{g} t}\right. \nonumber\\
&  \hspace{5em}\left.  \times\sum_{m=-2}^{2} \boldsymbol{\mathcal{G}}%
_{2m}(\theta,\varphi;f)F_{A}^{m}(\Omega)\right\}  .
\label{eq:app_random_real_displacement}%
\end{align}
Thus the sky and polarization components of the random incident field drive
the same lunar normal modes, while each frequency is weighted by the
corresponding transfer functions.

\subsubsection{Statistical average and spectral normalization}

Equation~(\ref{anisotropic_psd}) is used only after the random response has
been obtained. Substituting Eq.~(\ref{eq:app_stochastic_surface_amplitude})
into the two-frequency displacement covariance introduces $\langle\widetilde
h_{A}(f,\Omega) \widetilde h_{A^{\prime}}^{*}(f^{\prime},\Omega^{\prime
})\rangle$. The polarization and sky delta functions in
Eq.~(\ref{anisotropic_psd}) reduce the two polarization sums to one and the
two sky integrals to one. The resulting angular factor is $\int d\Omega
\,P(\Omega)\sum_{A}F_{A}^{m}F_{A}^{m^{\prime}\ast}=C_{mm^{\prime}}$, as in
Eq.~(\ref{eq:quadrupolar_covariance}), and the remaining frequency delta gives
Eq.~(\ref{eq:displacement_covariance}). There is no additional
polarization-angle average at this step.

The $\sqrt{2}$ in Eqs.~(\ref{eq:app_random_field}) and
(\ref{eq:app_random_real_displacement}) fixes the relation between this
complex amplitude and the one-sided spectrum of Eq.~(\ref{anisotropic_psd}).
At a fixed surface point, let $\boldsymbol{z}(t)=\int_{0}^{\infty
}df\,e^{i\omega_{g}t}\widetilde{\boldsymbol{u}}(f)$. For a stationary
background without a zero-frequency component, the nonconjugated correlations
have support at $f+f^{\prime}=0$ and vanish for $f,f^{\prime}>0$. Hence,
$\langle\boldsymbol{z}\cdot\boldsymbol{z}\rangle=0$ and
\begin{align}
\langle|\boldsymbol{u}|^{2}\rangle &  =2\langle|\Re\boldsymbol{z}|^{2}%
\rangle=\langle\boldsymbol{z}\cdot\boldsymbol{z}^{\ast}\rangle\nonumber\\
&  =\int_{0}^{\infty}df\,S_{h}(f)\sum_{m,m^{\prime}=-2}^{2}%
\boldsymbol{\mathcal{G}}_{2m}\cdot\boldsymbol{\mathcal{G}}_{2m^{\prime}}%
^{\ast}C_{mm^{\prime}}.\label{eq:app_variance_normalization}%
\end{align}
In the last line the kernels are evaluated at the same frequency and surface
point; the frequency delta has removed one frequency integral. This recovers
Eq.~(\ref{eq:total_mean_square}). Without the $\sqrt{2}$, taking the real part
would leave a factor $1/2$ for the amplitude covariance specified in
Eq.~(\ref{anisotropic_psd}). The factor is therefore part of the spectral
convention, not an additional enhancement of the physical response.

\subsubsection{Deterministic single-wave limit}

The same response operator also gives Eq.~(\ref{eq:coh_real_displacement}) for
a specified monochromatic plane wave. For frequency $f_{0}>0$, direction
$\Omega_{0}$, and real polarization coefficients $c_{A}$, take the
deterministic spectral input
\begin{equation}
\widetilde h_{A}(f,\Omega)=\frac{h_{0}}{\sqrt2}c_{A}\, \delta(f-f_{0}%
)\delta^{2}(\Omega,\Omega_{0}),\qquad\epsilon_{ij}=\sum_{A}c_{A}e^{A}%
_{ij}(\Omega_{0}). \label{eq:app_single_wave_input}%
\end{equation}
The delta functions specify a single frequency and direction; this input is
not assigned the stochastic covariance of Eq.~(\ref{anisotropic_psd}). In the
polarization convention of Appendix~\ref{app:coordinates}, $c_{+}%
=\mathcal{A}_{\nu_{0}}$ and $c_{\times}=\mathcal{B}_{\nu_{0}}$, so that
$\sum_{A}c_{A}F_{A}^{m}(\Omega_{0}) =\epsilon_{ij}\Lambda_{ij}^{(m)}%
=F^{m}(\Omega_{0},\nu_{0})$. Inserting Eq.~(\ref{eq:app_single_wave_input})
into Eq.~(\ref{eq:app_random_real_displacement}) removes the sky and frequency
integrals. The $1/\sqrt2$ in the input cancels the factor in the real-field
reconstruction, giving
\begin{equation}
\boldsymbol{u}_{\mathrm{coh}}(\theta,\varphi,t) =\Re\left\{  h_{0}e^{i2\pi
f_{0}t}\sum_{m=-2}^{2} \boldsymbol{\mathcal{G}}_{2m}(\theta,\varphi;f_{0})
F^{m}(\Omega_{0},\nu_{0})\right\}  . \label{eq:app_single_wave_recovery}%
\end{equation}
This is Eq.~(\ref{eq:coh_real_displacement}). It is obtained by specifying the
incident field, whereas Eq.~(\ref{eq:stochastic_response_amplitude}) retains
the full set of random incident amplitudes. Neither expression requires an
ensemble average to solve the linear response problem.

\end{document}